\documentclass[10pt,preprint]{aastex}
 
\usepackage{epsfig}
\usepackage{longtable}
\usepackage{graphicx}

\newcommand{\n}{\nodata}

\def\bi{\begin{itemize}}
\def\ei{\end{itemize}}
\def\be{\begin{equation}}
\def\ee{\end{equation}}

\def\gtrsim{\mathrel{\hbox{\rlap{\hbox{\lower4pt\hbox{$\sim$}}}\hbox{$>$}}}}
\def\lesssim{\mathrel{\hbox{\rlap{\hbox{\lower4pt\hbox{$\sim$}}}\hbox{$<$}}}}
\def\gtrsim{\mathrel{\hbox{\rlap{\hbox{\lower4pt\hbox{$\sim$}}}\hbox{$>$}}}}
\def\lesssim{\mathrel{\hbox{\rlap{\hbox{\lower4pt\hbox{$\sim$}}}\hbox{$<$}}}}

\slugcomment{Submitted for publication in the {\it Astronomical Journal}.}
\shortauthors{Lister and Homan}
\shorttitle{MOJAVE - I. VLBA 15 GHz Images}
\begin{document}

\title{MOJAVE: Monitoring of Jets in AGN with VLBA Experiments $-$
I. First-Epoch 15 GHz Linear Polarization Images}
 
\author{M. L. Lister}
\affil{Department of Physics, Purdue University, 525 Northwestern
Avenue, West Lafayette, IN 47907}
\email{mlister@physics.purdue.edu}
 
\author{D. C. Homan}
\affil{Department of Physics and Astronomy, Denison University,
Granville, OH 43023}
\email{homand@denison.edu}

\begin{abstract}

We present first-epoch, milliarcsecond-scale linear polarization
images at 15 GHz of 133 jets associated with active galactic nuclei
(AGN) from the MOJAVE survey. MOJAVE (Monitoring of Jets in Active
Galactic Nuclei with VLBA Experiments) is a long-term observational
program to study the structure and evolution of relativistic outflows
in AGN.  The sample consists of all known AGN with galactic latitude
$|b| > 2.5\arcdeg$, J2000 declination greater than $-20\arcdeg$, and
correlated 15 GHz VLBA flux density exceeding 1.5 Jy (2 Jy for sources
below the celestial equator) at any epoch during the period
1994$-$2003. Of the 133 AGN that satisfy these criteria, 96 are also
part of the VLBA 2 cm Survey.  Because of strong selection effects,
the sample primarily consists of blazars with parsec-scale
morphologies consisting of a bright core component at the extreme end
of a one-sided jet.  At least one third of the cores are completely
unresolved on the longest VLBA baselines, indicating brightness
temperatures above $10^{11}$ K. These cores tend to have electric
vectors that are better aligned with the inner jet direction, possibly
indicating the presence of a stationary shock near the base of the
jet. The linear polarization levels of the cores are
generally low ($<5\%$), but many of the extended jet regions display
exceedingly high fractional polarizations ($> 50\%$) and electric
vectors aligned with the jet ridge line, consistent with optically
thin emission from transverse shocks.  The cores and jets of the radio
galaxies show very little or no linear polarization.  Both the weak-
and strong-lined blazar classes (BL Lacs and high polarization radio
quasars) show a general increase in fractional polarization with
distance down the jet, but the BL Lac jets are generally more
polarized and have electric vectors preferentially aligned with the
local jet direction. We show that these differences are intrinsic to
the jets, and not the result of observational biases. We find that
distinct features in the jets of gamma-ray loud (EGRET) blazars are
typically twice as luminous as those in non-EGRET blazars, and are
more highly linearly polarized.  These differences can be adequately
explained if gamma-ray blazars have higher Doppler boosting factors,
as the result of better alignment with the line of sight and/or higher
bulk Lorentz factors.

\end{abstract}
 
\ 
\ 
\vfill\eject
\section{INTRODUCTION}

In terms of total energy output, active galactic nuclei (AGN)
represent some of the most energetic phenomena known in the universe,
and provide unique laboratories for studying a variety of high-energy
physical processes associated with special and general relativity,
particle acceleration, and magnetohydrodynamics. Of
particular interest are the roughly 10\% of all AGN that generate
massive outflows which produce powerful broad-band synchrotron and
inverse-Compton radiation. Direct imaging of the radio synchrotron
emission from AGN outflows using the Very Long Baseline Array (VLBA)
has revealed that they are highly collimated, relativistic jets of
material, with apparent superluminal speeds up to $\sim 40$~c
\citep{JMM01, KL04}. Images on larger scales show that many are able
to remain collimated for hundreds of kiloparsecs before depositing
enormous energies ($\sim 10^{61}$ ergs) into large plumes of
radio-emitting plasma.

The highly relativistic speeds of AGN jets have a substantial impact
on their observed properties, including their apparent speeds,
variability timescales, viewing angles, and apparent luminosities.
According to a stationary observer, the radiation from a relativistic
bulk flow will be highly Doppler-boosted (i.e., beamed) in a forward
direction. Any flux-limited sample selected on the basis of this jet
emission will therefore preferentially contain jets viewed nearly
end-on. For such an orientation, transverse speeds faster than light
are possible, and early VLBI observations (e.g.,\citealt{C77}) were
able to detect these apparent superluminal motions in bright, discrete
knots that moved outward at steady speed from a stationary ``core'',
assumed to be associated with an optically thick region near the base
of the jet.

The NRAO's\footnote{The National Radio Astronomy Observatory is a
facility of the National Science Foundation operated under cooperative
agreement by Associated Universities, Inc.} Very Long Baseline Array
(VLBA) has radically changed our view of AGN jets, making it possible
to routinely obtain high-dynamic range, full polarization images at
regular intervals. It is now evident that AGN jets are complex
magnetohydrodynamic phenomena that cannot be adequately modeled using
simple analytical methods or models involving purely ballistic
``knots''. They are best described as highly magnetized fluids that
respond to changes in the conditions at their nozzles and in
their external environments. The unsuitability of analytical models
implies that computer simulations are essential for fully
understanding the physics of jet phenomena. Indeed, there are many
groups that making important advances in this area (e.g.,
\citealt{HMD02,KSK02, AMG03}). However, to be effectively applied,
these simulations require a large set of observational constraints
that are best provided by VLBI arrays, because of their high angular
resolution and ability to penetrate through obscuring material in the
host galaxies. Although numerous detailed VLBI studies have been made
of nearby and unusual individual jets, so far there have been few
large statistical studies to investigate the overall properties of the
general jet population.  In an earlier study using the
Pearson-Readhead survey, we demonstrated the advantages in gathering a
large amount of data on a small sample \citep{LTP01}, by identifying
several new observational correlations and providing conclusive
evidence of relativistic beaming in blazars. However, little could be
inferred about the general jet population, because of the small sample
size (32 objects) and the presence of complex selection biases. These
biases can create artificial correlations in blazar samples and lead to
incorrect conclusions. Fortunately, it is possible
to properly account for these biases, given a complete sample that is
sufficiently large ($\gtrsim 100$ jets), with rigorous selection
criteria based on beamed jet emission \citep{LM97}.

In order to address these issues, we have begun a large, full
polarization VLBA campaign on a complete AGN sample with the goal of
providing the best possible dataset for investigating the statistical
properties of AGN jets and modeling their selection biases. MOJAVE,
which stands for Monitoring of Jets in AGN with VLBA Experiments, is a
complement to the recent VLBA 2 cm Survey (\citealt*{K98};
\citealt*{Z02}; \citealt*{KL04}). The latter program regularly imaged
over 150 AGN jets from 1994 to 2002 and made significant discoveries
regarding the general kinematics of AGN jets, including the finding
that highly superluminal jets are extremely rare, and that the
observed pattern speeds are directly connected with the underlying
flow. The 2 cm Survey did not, however, investigate the magnetic field
evolution of jets, since its observations were of total intensity
(Stokes I) only. Also, the survey was drawn from a low-frequency
all-sky catalog, and was statistically incomplete from the standpoint
of beamed jet emission. As we describe in
\S~\ref{sampselect}, MOJAVE uses a complete sample selected on the
basis of 15 GHz VLBI flux density only.

Although the 2 cm Survey contained the longest continuous monitoring
baselines of any VLBI survey carried out to date, reliable speeds
could only be measured in roughly $75\%$ of the sources
\citep{KL04}. In the remainder, the angular motions were either too
slow (because of low intrinsic speed or high redshift of the host
galaxy) or exhibited too few distinct features that could be tracked
during the observation interval. Also, a small number of nearby
sources displayed very fast motions that were under-sampled by the
monitoring intervals. The MOJAVE program is correcting these
deficiencies by increasing the monitoring baseline for the slow
sources by an additional three years, and sampling the faster sources
more frequently. With our current program, we observe each source at
intervals ranging from a few months to one year. We will have obtained
at least four additional epochs on each source by the end of our
program.  All of our calibrated data and images are available online
at the MOJAVE program
homepage\footnote{http://www.physics.purdue.edu/astro/MOJAVE}.

As MOJAVE is the first multi-epoch VLBI project to provide full linear
and circular polarization information on a large sample of AGN jets,
many of its goals are statistical in nature. Here we discuss the
overall linear polarization characteristics of our sample based on our
first-epoch observations only. Details on unusual individual sources,
as well as the overall evolution and circular polarization properties
of the sample will be presented in subsequent papers.

The overall layout is as follows. In \S~2 we discuss the selection
criteria and general properties of the MOJAVE sample. In \S~3 we
describe our observations and data reduction technique, as well as the
general method for deriving observational quantities. We discuss the
overall polarization characteristics of the sample in
\S~4, and describe distinct differences in the jets of BL Lac and
quasars, as well as gamma-ray and non-gamma-ray-detected blazars. We
summarize our main conclusions in \S~5. 

We adopt a contemporary cosmology with $\Omega_m = 0.3$,
$\Omega_\Lambda = 0.7$ and $H_o = 70 \; \mathrm{km\; s^{-1} \;
Mpc^{-1}}$. Our position angles are given in degrees east of north,
and all fractional polarization quantities refer to linear
polarization only. We use the convention $S_\nu \propto \nu^{\alpha}$
for the spectral index.

\section{THE MOJAVE SAMPLE}
\subsection{Selection criteria\label{sampselect}}

The goals of the MOJAVE program demand a statistically complete AGN
jet sample that is bright enough for direct fringe detection with the
VLBA, and whose properties can be compared with theoretical models
incorporating relativistic beaming effects (e.g.,
\citealt*{LM97}). Samples selected via total (single-dish) radio flux
density without regard to spectral index are not suitable since they
are dominated by radio galaxies with luminous, extended lobe
structure. Their parsec-scale jets are very faint, and imaging them
usually requires time-intensive VLBI phase-referencing
techniques. Since their jets lie much closer to the plane of the sky
than blazars, there is very little Doppler time compression, resulting
in much slower evolution and subluminal speeds. Also, simulating the
observational biases in such a sample requires detailed knowledge of
the relationship between intrinsic jet luminosity and extended lobe
power, which is currently not well-constrained.

An approach more suited to our goals involves constructing an AGN
sample based on beamed jet emission only. By using the milliarcsecond
scale (VLBA) 15 GHz flux density as a selection criterion, one can
effectively exclude the contribution from large-scale emission,
leaving a sample consisting almost entirely of radio loud AGN with
relativistic jets pointed nearly directly at us (i.e., blazars). The
only exceptions are a few nearby radio galaxies and several
peaked-spectrum sources whose jet axes likely lie much closer to the
plane of the sky.

The selection criteria we have chosen for the MOJAVE sample are
similar to those of the VLBA 2 cm Survey, and are summarized as
follows:
\begin{itemize}
\item{} J2000 declination  $\ge -20 \arcdeg$
\item{} Galactic latitude $|b| \ge 2.5 \arcdeg$
\item{} Total 15 GHz VLBA flux density of at least $1.5$ Jy ($\ge 2$
Jy for sources below the celestial equator) {\it at any epoch} during
the period 1994--2003.
\end{itemize}

Because of the variable nature of strong, compact AGN, we have not
limited the flux density criterion to a single fixed epoch. Doing so
would have excluded many highly variable sources from the sample, and
would subsequently reduce the robustness of statistical tests on
source properties.

\subsection{Selection method\label{selmethod}}
We have identified a total of 133 AGN that satisfy our selection
criteria (see Table 1), 96 of which are also members of the VLBA 2 cm
Survey. The procedure used in arriving at this list was somewhat
complex because of the limited VLBA monitoring data that were
available.  Our first step was to compile a candidate list of all
known AGN with an observed or extrapolated 15 GHz {\it single-dish}
flux density greater than 1.5/2 Jy. We consulted the K\"uhr 1 Jy
catalog \citep{KWP81,SMK94}, the VLA calibrator
database\footnote{http://www.vla.nrao.edu/astro/}, the Parkes
Half-Jansky flat spectrum sample \citep{DWF97}, and the JVAS survey
\citep{BWJ03} for this purpose. The latter is complete to within $
2.5\arcdeg$ of the galactic plane for northern declinations. Since
these combined catalogs are complete over our defined sky region at 5
GHz, the only potentially missing sources at this stage were those AGN
with spectra peaked at frequencies higher than 5 GHz, and variable
sources that happened to be weak at the time of the catalog selection.
To alleviate the former bias, we checked the WMAP catalog
\citep{BHH03}, the 22 GHz VLBI survey of \cite{MFP96}, and the
high-frequency peaked samples of \cite{TUW01} and \cite{DSC00} for
sources that met the single-dish flux density criteria. We also added
any variable sources from the UMRAO and RATAN AGN monitoring programs
\citep{AAH03,KKN02} that had sufficient 15 GHz flux density at any
epoch during the nine-year window. The latter program has observed
approximately 3000 radio sources with flux densities exceeding 100 mJy
since 1995. 

The second stage of source selection was to eliminate from the
candidate list all extended sources whose correlated VLBA flux
densities were weaker than our cutoff. Past surveys (e.g., the CJF
survey, \citealt*{TVR96}) have accomplished this by using a spectral
flatness criterion, which assumes that the weak-core sources will have
steep radio spectra because of their extended structure. We did not
adopt this method for our survey, since we found that some nearby AGN
with steep spectra but sufficiently bright cores (e.g., M87, 3C~380)
were incorrectly rejected. We instead used the measured VLBA flux
densities from the VLBA 2 cm Survey as the main criterion for MOJAVE
survey membership. If a candidate source still did not meet the VLBI
flux density criterion at four different epochs within the time window
(or had fewer than four VLBA epochs), we determined the VLBI-to-single
dish flux density ratio using near-simultaneous UMRAO and VLBA
observations (tabulated in \citealt*{K05}). We thus used this to
estimate the VLBI 15 GHz flux density based on RATAN or UMRAO
observations taken during the nine-year window.

\subsection{General properties\label{sampprops}}

In Table 1 we summarize the general properties of the MOJAVE
sample. The optical identifications in column (6) are from
\cite{VCV03}, with exceptions as noted. The BL Lac objects that
\cite{VCV03} have reclassified as quasars have on occasion displayed
emission lines slightly wider than the nominal 5 \AA $\;$ limit. Since
these are still much narrower than those typically found in quasars,
we retain the BL Lac classifications for these objects. We list the
optical V magnitude from \cite{VCV03} in column (7), and point out
that these are highly variable with time. The general
descriptions of the radio spectral shape in column (8) are from
\cite{KL04}. All but eleven sources have flat overall radio
spectra, which we define as a spectral index flatter than $-0.5$ at
any frequency between 0.6 and 22 GHz. The five steep-spectrum sources
in the sample have strong extended emission on arcsecond scales that
dominates the integrated spectrum, but their parsec-scale emission
still meets our selection criteria. The six peaked-spectrum
sources have no apparent radio emission on arcsecond scales (with the
exception of 2134+004; \citealt*{SOD98}). We do not include 
0738+313 and 1127$-$145 in the peaked-spectrum category, as they have
variable radio spectra that are occasionally flat \citep{KKN02}.

In column (9) of Table 1 we list the strongest measured 15 GHz VLBA
flux density during the period 1994--2003. Those values with asterisks
are estimated from single-dish measurements as described in
\S~\ref{selmethod}. Column (10) indicates whether the AGN has been
identified as a likely counterpart to a $\gtrsim 30$ MeV gamma-ray
source detected by the EGRET instrument on board the Compton Gamma-Ray
Observatory.  A ``Y'' indicates a highly probable EGRET identification
according to \cite{MHR01}, \cite{SRM03} and
\cite{SRM04}. Hereafter we will refer to these as
``EGRET sources''. A ``P'' indicates a less probable identification
according to these authors, and we will refer to these as ``probable
EGRET sources''. There are some disagreements in the literature
regarding these probable EGRET sources, with 0234$+$285, 1156$+$295,
and 2230$+$114, being considered as EGRET sources by
\cite{MHR01}, while 0446$+$112 , 0529$+$483 and 1936$-$155 are
considered as EGRET sources by \cite{SRM03,SRM04}. \cite{MHR01} do not
consider 0805$-$077, 1127$-$145, 1324$+$224, 1417$+$385 and 1504$-$167
to be probable EGRET sources. 

An overall summary of the optical and EGRET identifications for the
MOJAVE sample is given in Table~2. Optical counterparts have been
identified for 93\% of the sample, and the redshift data are currently
86\% complete.  The sample is dominated by quasars, with the
weak-lined BL Lacertae objects and radio-galaxies making up 17\% and
6\% of the sample, respectively.

\section{VLBA OBSERVATIONS AND DATA REDUCTION}

\subsection{Data calibration}
The first epoch of MOJAVE observations took place on 2002 May 31 using
the Very Long Baseline Array.  In this paper we present data obtained
from this and ten subsequent 24-hour MOJAVE sessions, up to and
including 2004 Feb 11. All 10 VLBA antennas were used at a central
observing frequency of 15.366 GHz. The data were recorded in eight
baseband channels (IFs) each of 8 MHz bandwidth using 1-bit sampling.
Both right and left hand polarizations were recorded simultaneously in
IF pairs, giving a total observing bandwidth of 32 MHz in each
polarization. During each observing session, short duration (roughly
seven minute) scans of 18 MOJAVE sources were interleaved in order to
span the widest possible range of hour angle, and to obtain maximum
parallactic angle and {\it (u,v)} plane coverage. The total
integration time on each source was approximately 65 minutes. There
was no need to observe any additional calibrators since all the target
sources were strong and compact.

The data were edited and calibrated at NRAO and Purdue University
following the standard methods described in the AIPS cookbook. The
individual IFs were kept separate throughout the processing, and were
only averaged together when making the final Stokes images. This was
necessary since the polarization leakage terms at each antenna usually
differ slightly among the IFs. A total of 80 leakage terms (four IFs
times two circular polarization hands, at ten antennas) were
determined at each epoch by running the task LPCAL on all sources, and
by taking the median solution values after removing obvious outliers.
We determined the relative instrumental electric vector position angle
(EVPA) rotation between epochs by comparing the individual leakage
term solutions over time \citep{GMA02}. With very few exceptions,
these remained extremely stable over periods of up to a year, and
based on the scatter we consider our relative EVPA error between
epochs to be less than two degrees. The leakage term method only gives
relative EVPA corrections between epochs, and still requires an
absolute correction to establish the EVPA directions on the sky. This
was obtained by comparing the VLBA EVPAs of the most compact MOJAVE
sources with single dish values obtained at the University of Michigan
Radio Observatory. The latter observations were carried out typically
within a week of the VLBA sessions. The systematic error in the
absolute EVPAs is dominated by the non-simultaneity of the
observations, and possible polarized structure not sampled by the
VLBA. Based on the single-dish comparisons, we estimate that our
absolute EVPA values are accurate to within five degrees.

\subsection{Imaging}
We used the Caltech Difmap package \citep{SPT94} to produce Stokes I,
Q and U images of each source from the calibrated visibility data
using natural weighting and a cell size of 0.1 milliarcseconds per
pixel. The best-fit elliptical Gaussian restoring beam FWHM dimensions
varied between 0.45 and 1.5 mas, depending on the declination of the
source.  The restoring beam parameters are listed in columns (3)--(5)
in Table 3. The typical rms noise level, obtained by examining pixels
well away from the source in each Stokes I image, was $\sim 0.2 \;
\mathrm{mJy\; beam^{-1}}$. Based on comparisons of highly compact
sources to near-simultaneous UMRAO observations, we estimate that the
absolute flux density scaling of our VLBA observations is accurate to
within 5\%.

In Figure~\ref{f:images} we display the first-epoch 15 GHz VLBA images
of the entire MOJAVE sample. Each panel contains two images of the
same source, the first consisting of I contours with linear fractional
polarization overlaid in color. The latter is defined as $P/I$, where
$P = \sqrt{Q^2 + U^2}$. The second image shows the single lowest I
contour plus contours of polarized intensity, and ticks indicating the
electric polarization vector directions. In most cases the lowest
contour levels correspond to three and five times the rms in I and P,
respectively. The linear scale factor of each image in parsecs per
milliarcsecond is given in column (12) of Table 3 for those sources
with a known redshift.

In columns (6) and (8) of Table 3 we list the total 15 GHz VLBA I and
P flux densities of each source. These are determined using the sums
of I, Q, and U intensities of all of the pixels shown in
Figure~\ref{f:images} whose I intensities exceeded three times the rms
noise level of the I image. Using
Rayleigh statistics, the noise in the P image was estimated to be 0.66
times the average of the Q and U rms noise levels. If the total P did
not exceed five times this value, the latter was used as an upper
limit in column (8). The integrated EVPA listed in column (9) is
calculated using the total Q and U flux densities of the source, where
$\rm{EVPA} = (1/2) \arctan{(U/Q)}$.

\begin{figure}
\epsscale{0.9}
\caption{\label{f:images}  (This figure is available at
http://www.physics.purdue.edu/astro/MOJAVE/paper1/fig1.ps.gz )
$\quad$ First-epoch 15 GHz VLBA images
of the MOJAVE AGN sample. Each panel contains two images of the same
source, the first consisting of I contours in successive integer
powers of two times the lowest contour (see Table 3), with
linear fractional polarization overlaid according to the color
wedge. A single negative I contour equal to the base contour is also
plotted.  The second image includes the lowest positive I contour from
the first image, and linearly polarized intensity contours, also in
increasing powers of two. The sticks indicate the electric
polarization vector directions.}
\end{figure}

\subsection{Visibility model fitting}

From observations of jet phenomena associated with young and evolved
stars in our galaxy, as well as nearby AGN jets such as those in M87
and Centaurus A, it is known that these outflows have highly complex
three-dimensional structures that do not easily lend themselves to
simple statistical analysis. However, the vast majority of AGN lie at
extremely large cosmological distances, and many of these details are
blurred out by the limited spatial resolution of current ground-based
radio interferometers. A close examination of the images in
Figure~\ref{f:images} shows several instances (e.g., 0007$+$106,
$0446+112$, 0642$+$449, 1751$+$288) where no actual jet is evident,
since it lies below the resolution and/or sensitivity level of the
VLBA at this frequency. At the other extreme, there are also cases
where the jet is clearly resolved in a transverse direction (e.g.,
0836+710, 1641+399, 1828+487), making a more detailed analysis
possible. For the statistical purposes of this paper, we have fit each
source with a minimum number of elliptical Gaussian ``components''
that when convolved with the restoring beam, generally reproduce the
observed structure. In a separate paper we will present a more
detailed ridge line-based analysis of approximately a third of the
sample members whose jets are not fully described by multiple Gaussian
fits.

We carried out our analysis in the {\it(u,v)} plane by fitting the
brightest features in each source with two-dimensional Gaussian
components. Using the ``modelfit'' task in DIFMAP, we fit several
components to each source, resorting to zero-size (delta-function)
points if the least-squares routine shrank the fitted Gaussian major
axis to extremely small values. These fits provided positions, sizes
and flux densities for the brightest, most compact jet features, which
are tabulated in Tables 4 and 5. Based on our previous 15 GHz studies (see
Appendix A of \citealt*{HOW02}), we estimate typical errors of 5\% on the
total intensities and positional errors of $\sim 1/5$ of the restoring
beam dimension. For the brightest components, the errors are smaller
by approximately a factor of two. 

The milliarcsecond morphology of most sources consists of a strong,
flat-spectrum ``core'' component that is typically unresolved along one
or both axes of the restoring beam and is located at the extreme end
of a more diffuse jet structure. This core component is generally
thought to represent the region near the base of the jet where it
becomes optically thick.  Observational evidence of apparent core
shifts with frequency (e.g., \citealt*{L98}) support this view. 

\subsubsection{Core measurements}

We list the position of the fitted core component in polar coordinates
with respect to the map center in columns (2) and (3) of Table 4. In
most cases the core offset is negligible (less than one pixel, or 0.1
milliarcsec), since the phase self-calibration process tends to shift
the map center to lie on the brightest feature in the
source. Exceptions arise when a strong feature lies very close to the
core (e.g., 0224+671), or when the source is dominated by an
exceptionally strong feature in the jet (e.g., 0923+392). In sources
with a non-zero core offset, there is also the possibility that we are
seeing both the jet and the counter-jet. In blazars the latter is not
usually visible since its radiation is beamed away from us. There are
four sources in our sample in which counter-jets have been detected in
other studies: 0316+413 (3C~84), 0238$-$084 (NGC 1052), 1413+135, and
1957+405 (Cygnus A). The exact location of the core in these sources
is not well-defined, because of possible foreground absorption in the
vicinity of the central nucleus. In such cases we have considered the
feature closest to the map center to be the core. Other sources in our
sample which may have visible counter-jets are 0742+103, 1548+056, and
2021+614. Multi-frequency imaging of these AGN would aid in
identifying the core based on its predicted flat spectral index, 
variability, and high brightness temperature.  Such a study of
2021+614 is currently underway (Lister et al., in preparation).

The fitted flux density and Gaussian parameters of the core components
are given in columns (4) through (7) of Table 4. The brightness
temperatures in column (8) are calculated in the observer frame
assuming a Gaussian profile using the formula $$T_b = { 1.221 \times
10^{12} \; S_{core} \over \nu_{obs}^2 \; a_{maj}\; a_{min}} \quad
\mathrm{K},$$ where $S_{core}$ is the core flux density in Jy,
$\nu_{obs}$ is the observing frequency in GHz and $a_{maj}$ and
$a_{min}$ are respectively the FWHM major and minor axes of the fitted
core Gaussian in milliarcseconds. It should be noted that these are
``best-fit'' brightness temperatures, and in many cases, may in fact
be lower limits because of the limited spatial resolution of the data.
For those cores which were completely unresolved along at least one
axis of the restoring beam (such that the best fit was a
delta-function), a lower limit on $T_b$ was calculated assuming a
minimal resolvable component size. The expression for the latter,
based on the signal-to-noise ratio at the core position and the
restoring beam dimensions, is given by \cite{K05}.

Columns (9) and (10) list the fractional polarization and EVPA of the
core, which are calculated from the mean I, Q, and U of the nine
contiguous pixels centered at the fitted core position. In cases where
the polarization is below the lowest P contour level listed in Table
2, upper limits are given corresponding to five times the P rms noise
level.

\subsubsection{Jet measurements}

We use the same Gaussian component method to describe the jet emission
in the MOJAVE sample. The results of these fits are given in Table
5. Additionally, we measure the mean jet polarization by integrating
over all jet emission.  This is determined in the image plane using
only those pixels that are above the base I contour level listed in
Table 1. We exclude pixels in the core region defined as
$$\left({x \over b_{min}}\right)^2 + \left({y \over b_{maj}}\right)^2
\le {\ln{\left|I_{core} / I_{base}\right|} \over 4 \ln{2}},$$
where $b_{maj}$ and $b_{min}$ are the FWHM major and minor axes of the
Gaussian restoring beam, and $x$ and $y$ are the right ascension and
declination coordinates of the pixel with respect to the
core. $I_{base}$ is the lowest contour level in the image and
$I_{core}$ is the pixel intensity at the core position. Outside this
region, the emission from the core is below the lowest contour
level. For most sources, it corresponds to an ellipse with dimensions
approximately equal to those of the Gaussian restoring beam listed in
Table 3.

\section{DISCUSSION}

\subsection{Core component properties}

Although the flat-spectrum core components are responsible for much of
the polarized and total flux density in our VLBA images, their
properties are the most difficult to interpret because they likely
represent a blend of many small individual emitting regions close to
the base of the jet. In roughly one third of the sources, the core
feature could be adequately fit by an unresolved point model, yielding
only a lower limit on the brightness temperature. The observed
brightness temperatures of the remaining sources are typically above
$10^{11}$ K, which is close to the detection limit of ground-based
interferometers for cores with flux densities of a few hundred
mJy. For this reason we expect that many of the $T_b$ values listed in
Table 4 are also lower limits.  Indeed, a full {(\it u,v}) plane
analysis of the fine-scale structure of 250 flat-spectrum AGN in the
MOJAVE and 2 cm Survey samples by \cite{K05} shows that in over 65\%
of the sources, at least half of the total flux density is unresolved
on the longest baselines. 

\begin{figure}
\plotone{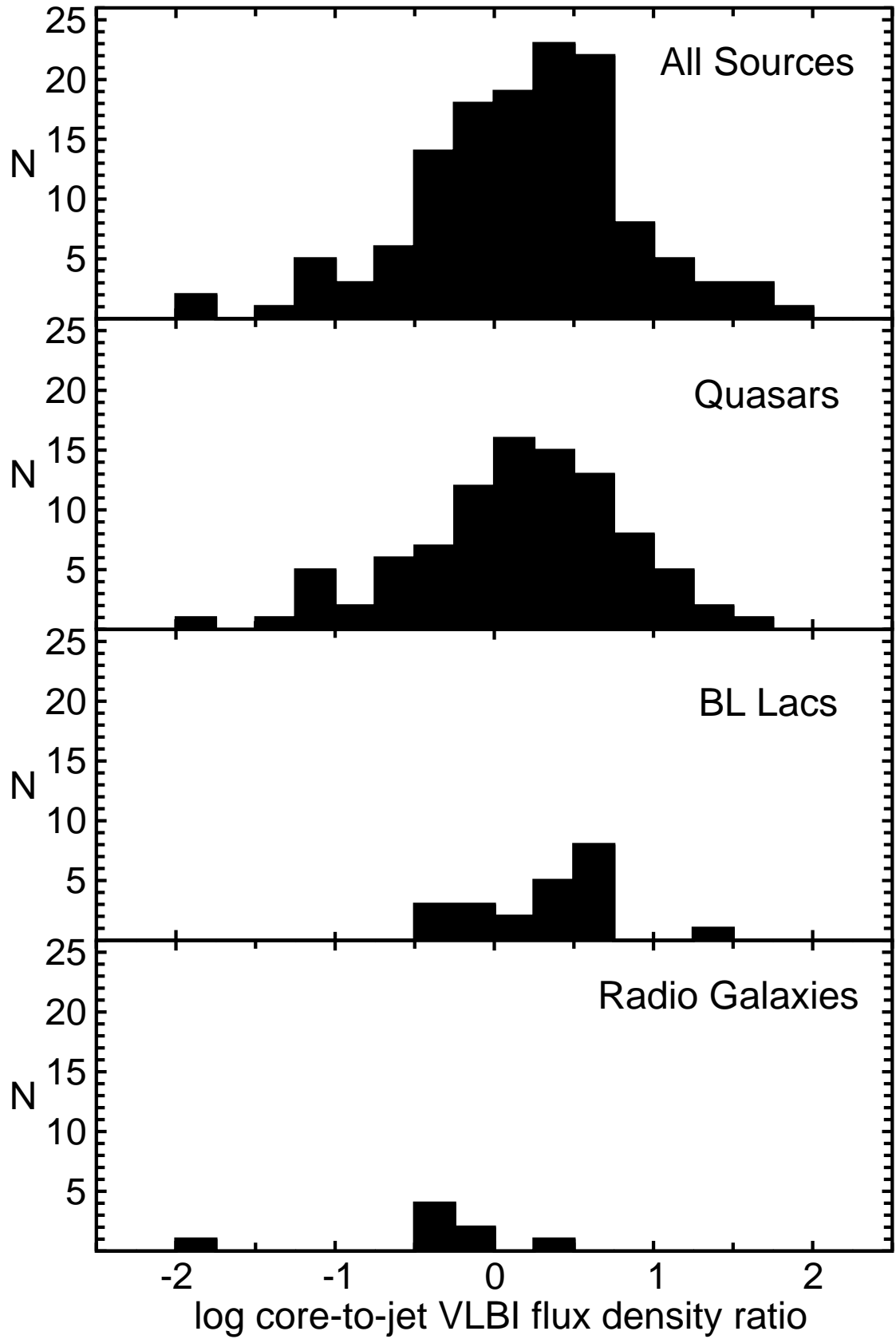}
\caption{\label{f:rj} Distribution of VLBI core-to-jet ratio ($R_j$)
for the full MOJAVE sample (top panel), and major optical subclasses
(lower panels). }
\end{figure}

The relative prominence of the core component with respect to the
remainder of the VLBI jet varies significantly among the sources in
our sample. We define the core-to-jet flux density ratio as $$R_j =
{S_{core} \over S_{jet} (1+z)^{0.7}},$$ where we have assumed typical
spectral indices of 0 and $-0.7$ for the core and jet,
respectively. For sources without measured redshifts we assume a
typical value of $z = 1$. The distribution of $R_j$
(Figure~\ref{f:rj}) is very broad, spanning four orders of magnitude,
with a peak at approximately $R_j = 3$. The spread in $R_j$ is likely
a result of strong flux variability in the core components, as well as
a wide range of Doppler factor in the jet population. The latter
broadens the distribution since the core and jet emission, having
different spectral indices, are expected to be beamed by different
amounts. We do not find $R_j$ to be correlated with either redshift or
angular size distance.

\begin{figure}
\plotone{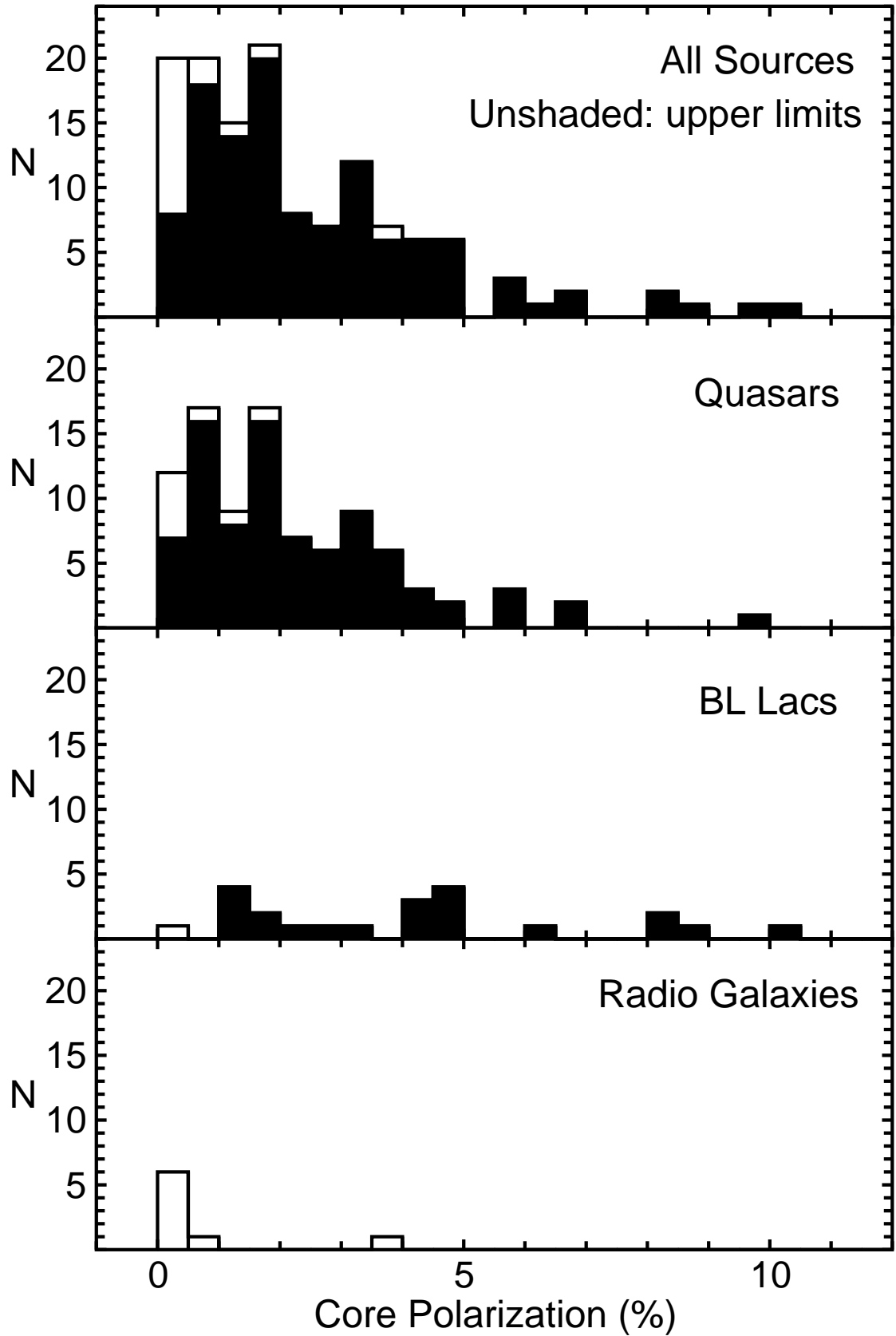}
\caption{\label{f:mcore} Linear fractional
polarization of the VLBI core component for the entire MOJAVE sample
(top panel) and three major optical subclasses (lower panels). The
unshaded boxes represent upper limits on those cores in which no
linear polarization was detected.  The distribution for the BL Lacs
differs from the quasars at 99.9\% confidence according to a Wilcoxon
test. } 
\end{figure}

The core components have typically low fractional polarizations
($m_{core} \lesssim 5\%$), and range up to a maximum of $\sim 10\%$ in
the case of $1538+149$ (Fig.~\ref{f:mcore}). No polarized emission was
detected in any of the eight radio galaxy cores. This was also the
case for seven quasars and one BL Lac object (1413+135).  As we 
discuss in \S~\ref{S:jetprops}, the fractional polarization levels are
generally much higher in jet regions downstream from the core.

\begin{figure}
\epsscale{.65}
\plotone{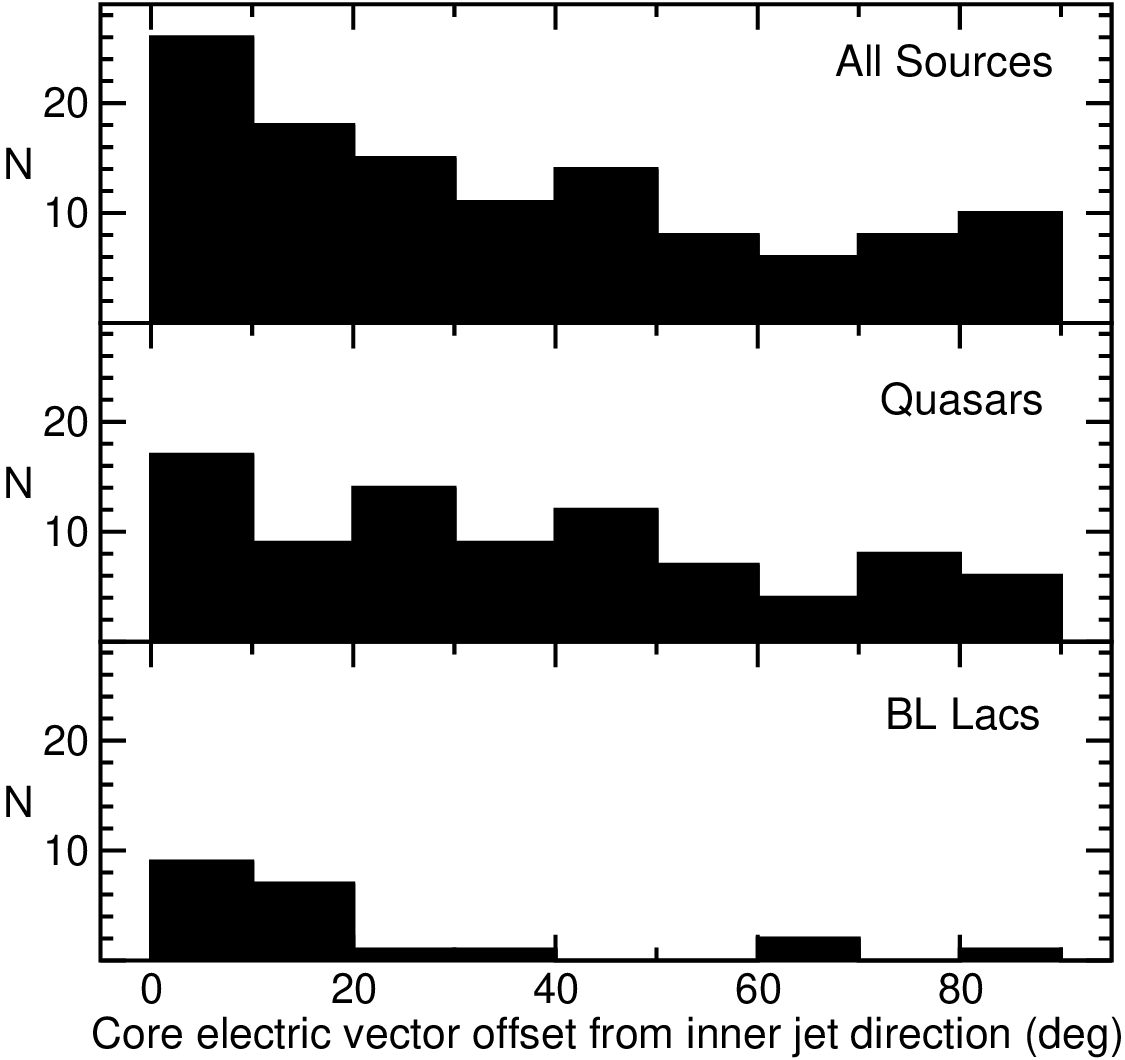}
\caption{\label{f:coreevpa} Distribution of electric vector offset
with respect to inner jet direction at the locations of VLBI cores in
the MOJAVE sample. The inner jet direction is defined as the position
angle of the Gaussian jet component closest to the core.  The BL Lac
and quasar distributions differ at the 99.9\% confidence level
according to a K-S test.  }
\end{figure}

The linearly polarized emission may be strongly depolarized in the
core regions, either from an external screen or internal plasma
processes. This would also imply a random rotation of the observed
electric vectors with respect to the jet, however, this is not the
case for the BL Lacertae cores (Figure~\ref{f:coreevpa}). The latter
tend to have core electric vectors that are well-aligned with the
inner jet direction. Faraday depolarization without accompanying
rotation is possible, but typically requires special geometries that
include a mixture of Faraday thin and thick material on scales smaller
than the restoring beam.  Some quasars are known to have large
parsec-scale rotation measures \citep{ZT04}, but there are currently
not enough data to indicate whether there are strong differences in
the Faraday screens of weak- and strong-lined blazars. If such
differences do exist, they must persist out to several hundred parsecs
out from the central engine, since we show in \S~\ref{BLQSOjet} that
these polarization differences are also present in jet regions
well downstream from the cores.

The case for Faraday depolarization is much stronger in the seven
sources in our sample for which very little or no polarized emission
is present in our images. The jets of the radio galaxies 0007+106,
NGC~1052, M87, Cygnus A, and 2021+614 lie at larger viewing angles,
and are heavily depolarized by material associated with an obscuring
torus located in their host galaxies. The two remaining unpolarized
sources, the BL Lac object 1413+135 \citep{PSC02} and the
peaked-spectrum quasar 0742+103 \citep{BPB03}, also display heavy
absorption features in their central regions. The nearby radio galaxy
3C~84 also suffers from heavy foreground absorption \citep{WDR00}, and
we detect polarized emission only in the southern lobe that is
presumably located closer to us (Figure 1).

\begin{figure}
\epsscale{.65}
\plotone{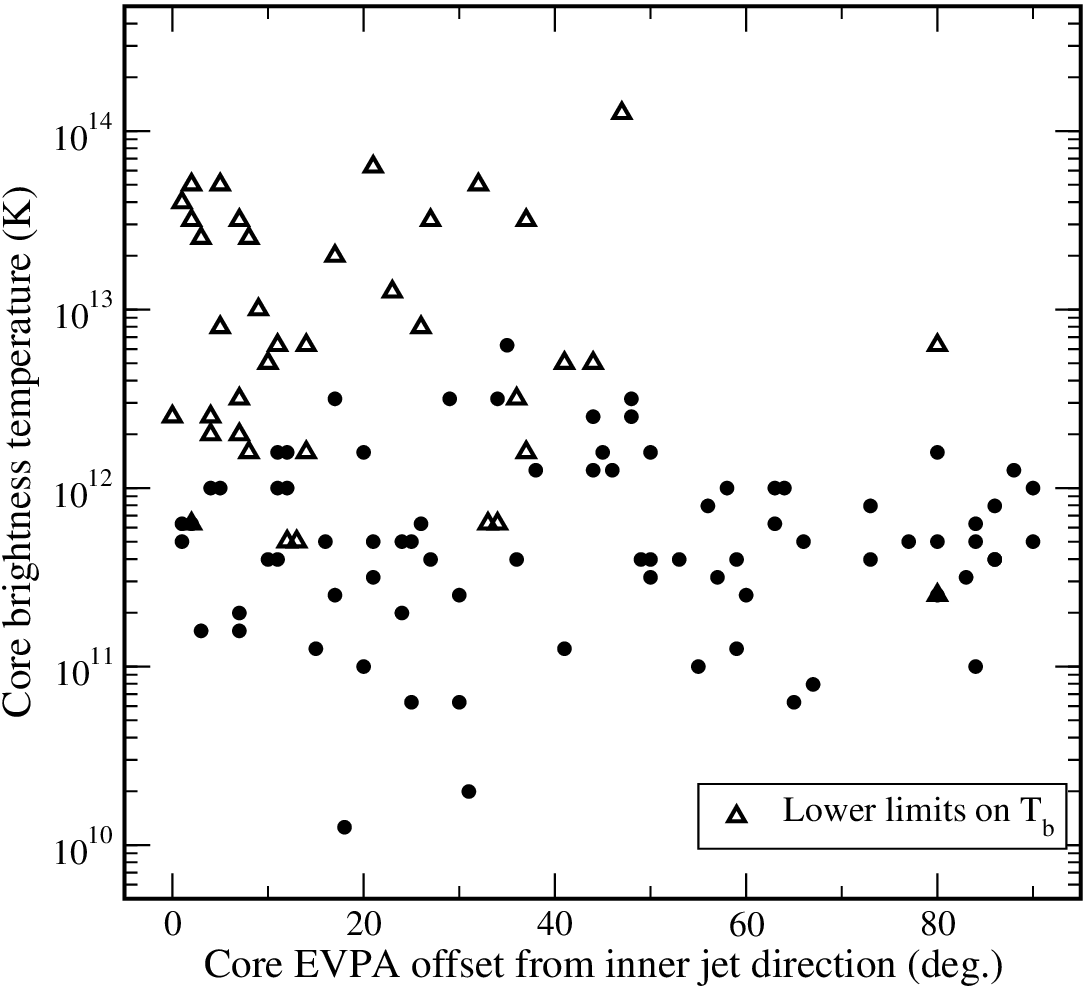}
\caption{\label{f:coreevpavstb} Observed brightness temperature
versus electric vector misalignment angle for core components in the
MOJAVE sample. The filled circles denote best-fit brightness
temperatures, and the open triangles denote lower limits. Cores with
the highest brightness temperatures tend to have electric vectors that
are better aligned with the inner jet direction. }
\end{figure}

Apart from possible Faraday depolarization, the cores may appear
weakly polarized because of optical depth effects. Synchrotron
radiation from an optically thick region such as the core will tend to
be weakly polarized because the fields are usually not well-ordered
over the short path length into the plasma \citep{H05}. Also, if the
core region contains regions of orthogonal electric vectors, these can
combine within the VLBI restoring beam to lower the net
polarization. We find a tendency for the most unresolved cores (those
with only lower limits on brightness temperature) to have electric
vectors that are better aligned with the inner jet direction
(Figure~\ref{f:coreevpavstb}).  Such an alignment might be expected,
for example, from a strong standing shock near the base of the
jet. The misaligned cases may simply be instances where we have
insufficient spatial resolution to separate one or more components
close to the base of the jet. Since such features will eventually
propagate downstream, it will be possible to investigate this scenario
more fully using the full multi-epoch MOJAVE polarization dataset.

\subsection{\label{S:jetprops}Jet properties}

At least 94\% of the MOJAVE sources have one-sided jet morphologies,
which is a strong reflection of the sample selection biases.  If the
milliarcsecond-scale emission from these jets is highly anisotropic
(i.e., beamed along the flow direction), the sources with the highest
flux densities will tend to have their jet axes preferentially aligned
toward the observer. Emission from the receding jet will be beamed in
the opposite direction, rendering it nearly undetectable. This
tendency toward one-sided jets is well-known from previous VLBI
surveys (see review by \citealt*{UP95}), and is considered to be key
evidence for relativistic bulk motion in AGN jets.

There are still two ways that lesser-aligned jets (lying closer to the
plane of the sky) can meet the MOJAVE selection criteria and yet
appear as two-sided. They may have a) sufficient flux density merely
because of their proximity (e.g., NGC~1052, Cyg~A, 3C~84), or b)
sufficiently small intrinsic size that their (isotropic) lobe emission
dominates their milliarcsecond-scale flux. Some of the peaked-spectrum
radio sources, believed to be young jets, fall into this
category. Possible examples in MOJAVE include 0742+103 and 2021+614.

\begin{figure}
\plotone{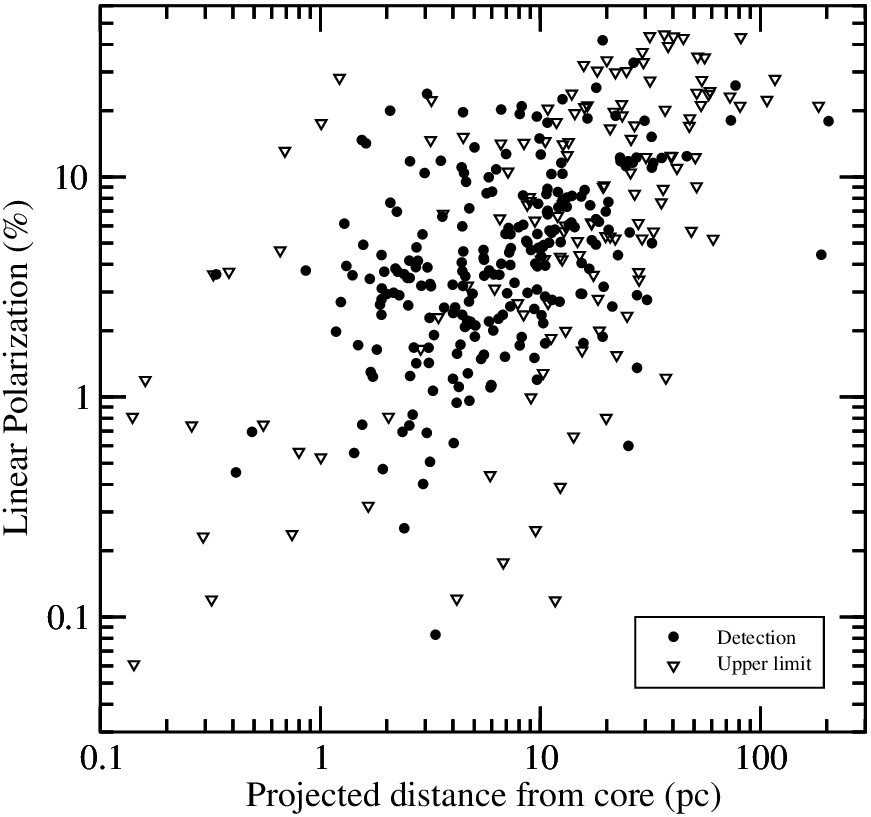}
\caption{\label{f:rmas_vs_mcpt_lim} Plot of linear fractional polarization of
individual jet components versus projected distance from the VLBI core
component. The filled circles indicate polarization detections, while
the open triangles represent upper limits. }
\end{figure}

Despite the large amount of foreshortening caused by the orientation
bias of the MOJAVE sample, we still find several major trends in the
observed jet properties. In Figure~\ref{f:rmas_vs_mcpt_lim} we plot
the linear fractional polarization of fitted jet components against
their projected distance from the core in parsecs.  Ignoring the
artificial trend of the upper limits getting larger with distance
(because of the weakening flux density of the jet components), there
is a clear upward trend of fractional polarization, which is confirmed
by Kendall's tau test on censored data at greater than 99.99\%
confidence. This increase in polarization with distance downstream has
also been seen in smaller samples at 5 GHz
\citep{CWR93}, 15 GHz \citep{HOW02}, 22 GHz \citep{LS00}, and 43 GHz
\citep{L01}.  The polarizations at the sites of the fitted Gaussians
range up to a maximum of $\sim 40\%$ seen in 3C~279 (1253$-$055), with
more typical values of $\sim 5\%$. However, in many jets the maximum
in fractional polarization does not occur at the locations of fitted
components, with some regions being more than 50\% linearly polarized
(Fig.~\ref{f:images}). Such high values indicate extremely
well-ordered fields in optically thin synchrotron-emitting regions,
and can potentially be used to constrain the viewing angle to the jet
(e.g., \citealt*{CW88}).

\begin{figure}
\plotone{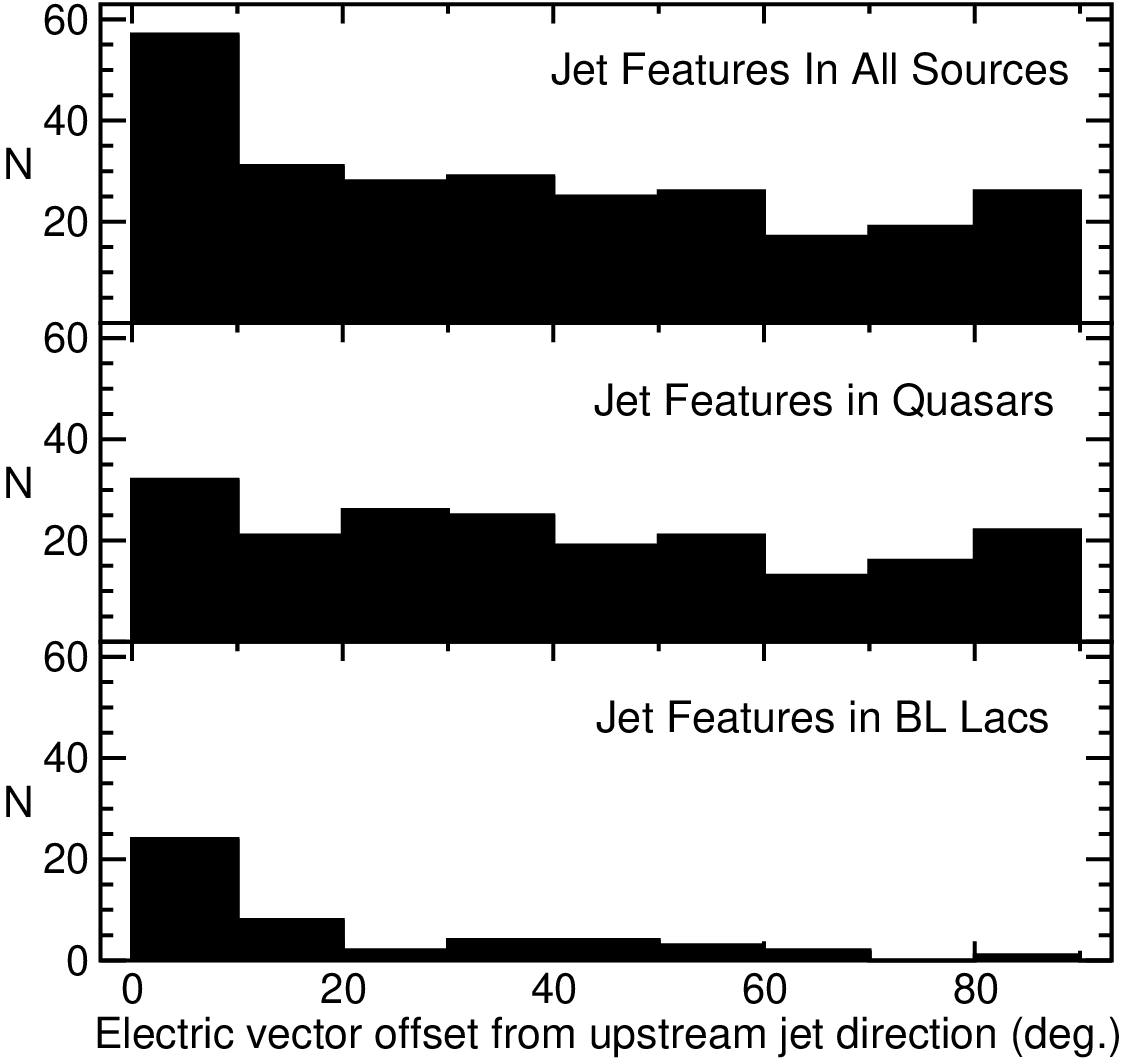}
\caption{\label{f:QSOBL_EVPAup} Distribution of electric vector offset
with respect to the upstream jet direction at the locations of jet
components in the full MOJAVE sample (top panel), and for optical
subclasses (lower panels). The upstream jet direction is defined as
the position angle to the nearest component located closer to the
core. The BL Lac and quasar distributions differ at greater than
99.99\% confidence according to a Wilcoxon test. }
\end{figure}

The EVPA orientations of the jet components lie at a variety of angles
to the local jet direction, which we define as the direction to the
nearest component upstream (Fig.~\ref{f:QSOBL_EVPAup}). The
distribution of EVPA offset is peaked at zero degrees,
indicating an excess of jet components with electric vectors aligned
with the jet. Assuming that the jet components are optically thin and
that the magnetic field lies perpendicular to the observed electric
field vectors, the form of the distribution is similar to that
predicted for an ensemble of oblique shocks with random orientations
within the jet \citep{LMG98}. Transverse shocks aligned with the flow
always have EVPA offsets of zero, regardless of observer viewing angle
\citep{CC90}, so these cannot be dominant in quasar jets. There is no
overall trend between the EVPA offset and projected distance from the
core as seen by \cite{LS00} at 43 GHz. Our observations are probing
regions much further downstream at poorer spatial resolution, however,
and the steep fall-off in jet intensity typically prevents polarized
emission from being detected in the outermost components.

\subsection{\label{optical}Connections with optical emission line properties}
There are now several studies (\citealt{CWR93,LMG98,L01}) that have
found substantial differences in the parsec-scale polarization
properties of weak- versus strong-lined blazars. These AGN classes are
commonly referred to as high-polarization radio quasars (HPRQ) and BL
Lacertae objects, respectively, with the division occurring at a
rest-frame equivalent emission line width of 5\AA. This number is
somewhat arbitrary, since historically there has been little evidence
for any bi-modality in the emission line width distribution of
radio-loud AGN, although \citealt*{LPP04} have proposed a new
classification scheme based on a bi-modality of O[III] line
width. Furthermore, the duty cycles of emission line variability are
not well known, making the BL Lac classification of individual objects
often controversial.

Given these caveats, we use the most prevalent BL Lac and quasar
classifications found in the literature (see Table 1) to re-examine
the possible differences in the parsec-scale jets of these two optical
classes using a large, complete blazar sample. For quantities that do
not involve limits, we use the Kolmogorov-Smirnov (K-S) test, and for
censored quantities we use Gehan's generalized Wilcoxon test from the
ASURV survival analysis package \citep{ASURV}.

\subsubsection{Core component properties}
We find the VLBI core components of BL Lacs to be typically more
polarized than the quasars (Fig.~\ref{f:mcore}).  They also have
electric vectors preferentially aligned with the inner jet direction,
whereas the quasar cores show no preferred E-vector orientation
(Fig.~\ref{f:coreevpa}).  Although the median core-to-jet ratio for BL
Lacs appears to be slightly higher than the quasars (Fig.~\ref{f:rj}),
the K-S test only gives a 89\% probability that the two are drawn from
different parent distributions.
 
\subsubsection{Jet properties\label{BLQSOjet}}
\begin{figure}
\plotone{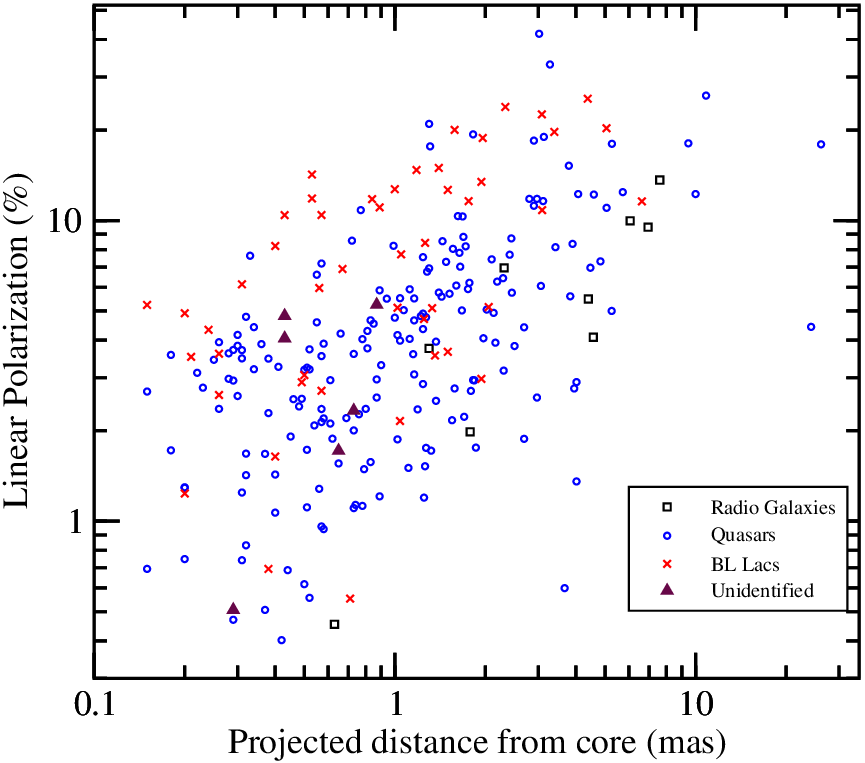}
\caption{\label{f:rmas_vs_mcpt} Plot of linear fractional polarization of
individual polarized jet components versus angular projected distance
from the VLBI core component. The open circles and squares denote
those associated with quasars and radio galaxies, respectively, while
the crosses denote those associated with BL Lac objects. The open
triangles belong to jets with no known optical counterparts. }
\end{figure}

\begin{figure}
\plotone{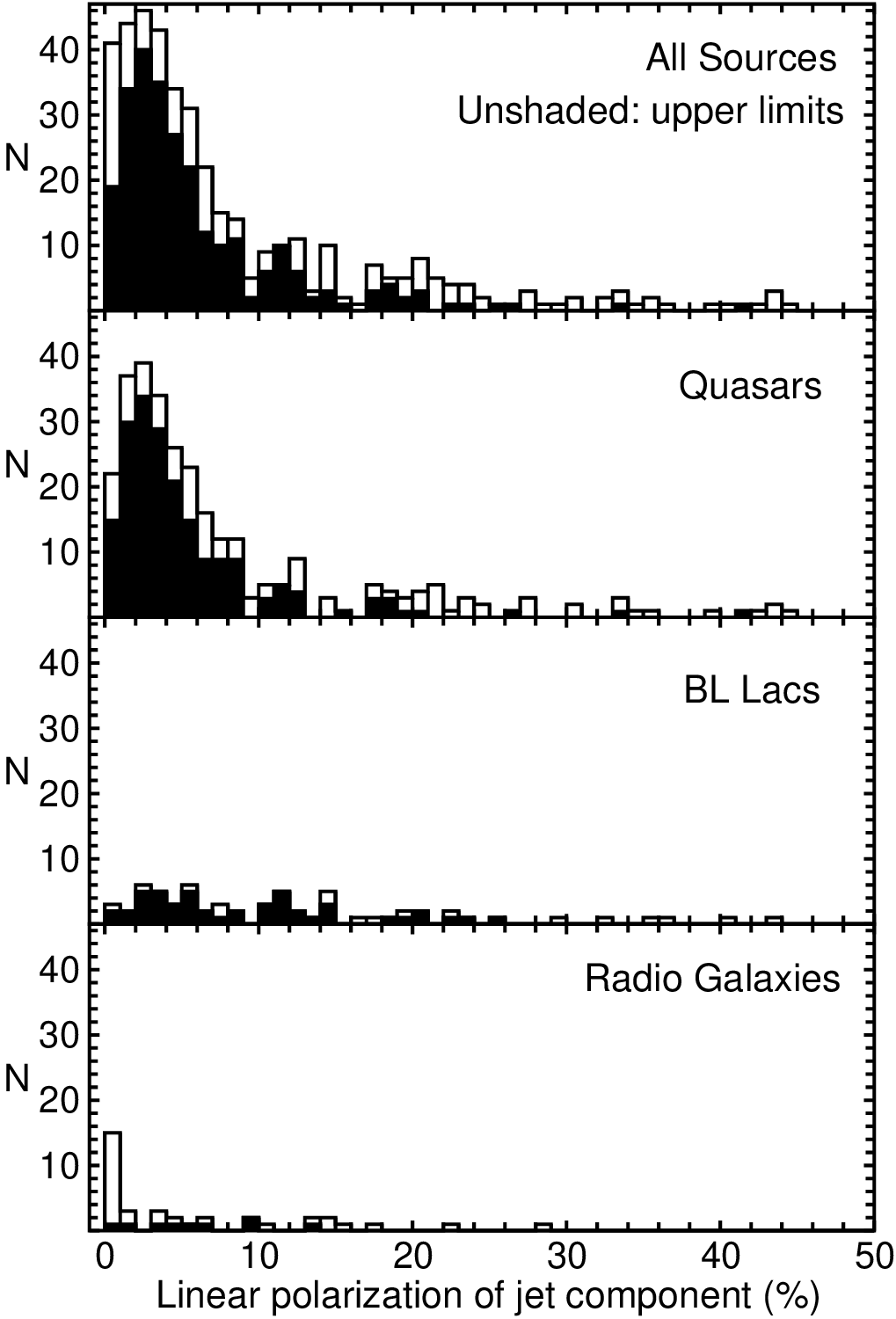}
\caption{\label{f:mcpt} Distributions of linear fractional
polarization for jet components in the full MOJAVE sample (top panel),
and optical sub-classes (lower panels). The unshaded boxes represent
upper limits for those components with no detected polarization. The
quasar and BL Lac distributions differ at greater than $99.99\%$
confidence according to a Wilcoxon test. }
\end{figure}

\cite{L01} found that at 43 GHz, the jets of BL Lac objects in the
Pearson-Readhead sample tend to be more polarized at a given projected
distance from the core than those of quasars. We confirm these
findings using the much larger MOJAVE sample at 15 GHz. At our lower
observing frequency, it is possible to image much more of the faint,
steep-spectrum regions located further down the jet, at the expense of
poorer spatial resolution. In Figure~\ref{f:rmas_vs_mcpt} we plot
fractional polarization versus projected angular distance from the
core for all jet components in the MOJAVE sample with detected
polarized emission. Although the BL Lac jet components are spread over
a large region of the diagram, most lie near the top of the
distribution, as also seen at 43 GHz by \cite{L01}. The jet components
of BL Lacs are not only more polarized than quasars
(Fig.~\ref{f:mcpt}) but also have electric vectors preferentially
aligned with the ridge line (Fig.~\ref{f:QSOBL_EVPAup}).

\subsubsection{Possible luminosity and redshift biases}

\begin{figure}
\epsscale{.95}
\plotone{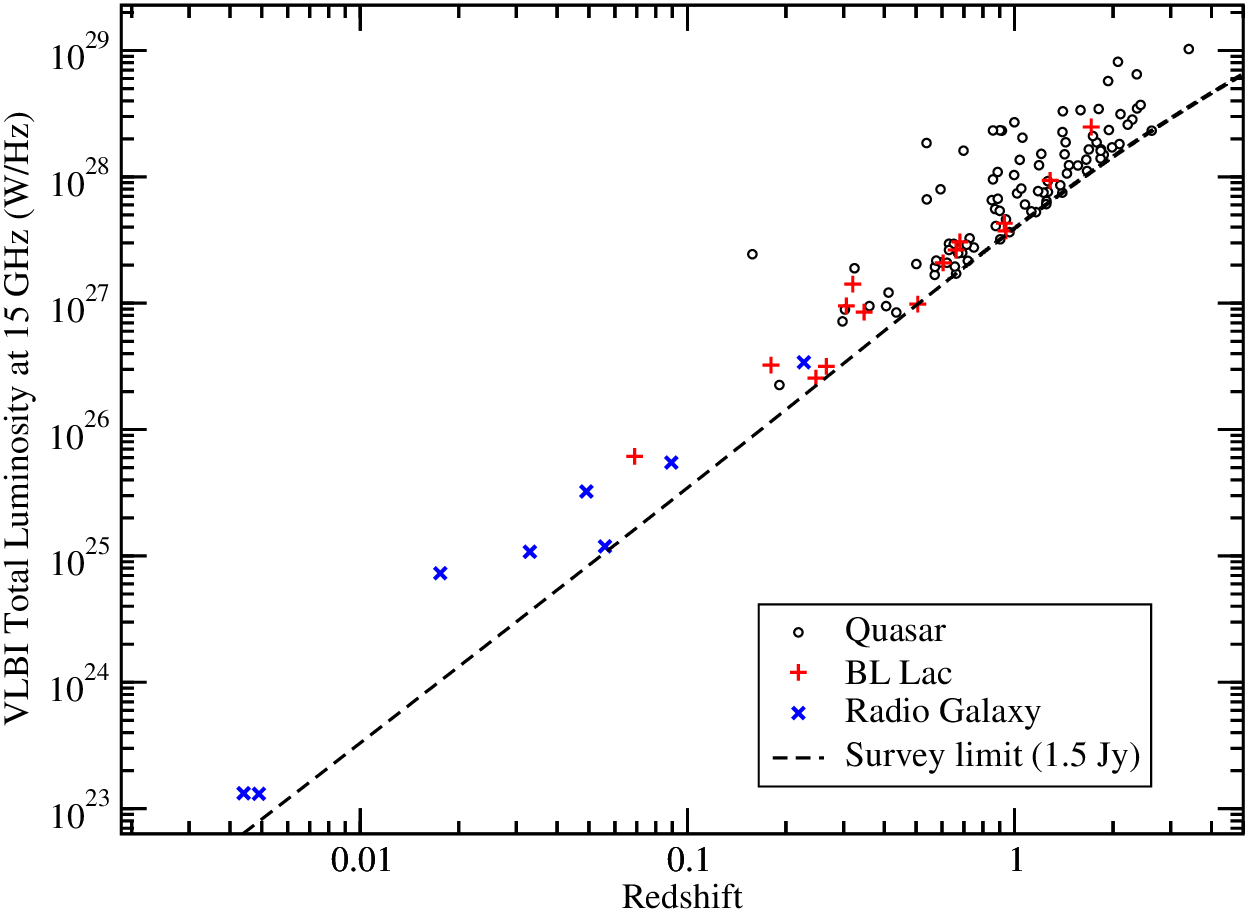}
\caption{\label{f:LvsZ} Maximum 15 GHz total VLBI luminosity versus
redshift for optically identified MOJAVE sources. The dashed line
corresponds to the survey limit of 1.5 Jy for northern sources,
assuming flat radio spectra. The open circles denote quasars, the plus
symbols denote BL Lac objects, and the x's represent radio galaxies.}
\end{figure}

One possible explanation for the polarization differences in quasars
and BL Lacs is that the latter are preferentially located at lower
redshifts, and therefore have lower luminosities in our flux-limited
sample. In Figure~\ref{f:LvsZ} we show a plot of maximum total 15 GHz
VLBI luminosity versus redshift for the optically identified objects
in the MOJAVE sample. K-S tests indicate that the BL Lacs are
concentrated in a different region of the L-z plane than the quasars.

As a check on possible biases in our analysis, we re-performed our
statistical tests on sub-samples that included only those BL Lacs and
quasars with maximum 15 GHz VLBI luminosity greater than $8 \times
10^{26} \; \mathrm{W \; Hz^{-1}}$. The BL Lacs without known redshifts
were also included. We selected a subsample of 48 quasars that closely
matched the $L-z$ region occupied by the BL Lacs, such that the two
redshift, angular size distance, and maximum 15 GHz VLBI luminosity
distributions were not statistically different according to a K-S
test. We verified that the core and jet luminosity distributions of
the two subsamples were also statistically indistinguishable.

We find that the differences in polarization between the two optical
classes are still present in these matched samples, with drops in
confidence level of less than 1\%, mainly as a result of the smaller
sample sizes. Therefore, the differences cannot be attributed to
luminosity bias. The presence of these apparent differences at both 15
and 43 GHz argues against Faraday screens being the culprit, unless
the rotation measures are exceptionally high. The fact that many
polarized jet regions with $m > 10$\% are seen in both BL Lacs and
quasars suggests that  this is unlikely.

Our observations therefore indicate that the magnetic field properties
of BL Lac and quasar jets are intrinsically different, either as a
result of internal properties, such as jet power or speed, or the
external environment through which the jet is propagating. There is
already evidence that BL Lac jets may be the beamed versions of
low-power FR-I type radio galaxies, while quasars are associated with
FR-II sources \citep{UP95}. Numerical simulations suggest that strong
shocks form more easily in low-Mach number flows (P. Hughes, private
communication). In this case, BL Lac jets may preferentially contain
strong, transverse shocks that increase the field order to a higher
degree than weaker, oblique shocks that develop more frequently in the
jets of quasars. Another possibility, raised by \cite{JML05}, is that
forward and reverse shocks may have fundamentally different
polarization properties, with the latter being more prevalent in BL
Lac objects.

\subsection{\label{gamray}Connections with gamma-ray properties}

During its years of operation between 1991 to 1995, the EGRET
instrument on board the Compton Observatory detected approximately 60
blazars at energies above $\sim 30$ MeV. There has been much
discussion as to whether these ``EGRET blazars'' are intrinsically
different in some respect from the general blazar
population. Gamma-ray emission models involving inverse-Compton
scattering (e.g., \citealt*{DSM92}) suggest that the gamma-rays are
highly relativistically beamed, perhaps much more so than the radio
synchrotron emission. It has been suggested that the blazars detected
by EGRET have particularly high Doppler factors \citep{LV03}, and
kinematic studies \citep{KL04} have indeed shown that they have faster
superluminal jet speeds. This simple picture is complicated, however,
by the highly variable nature of AGN at all wavelengths, particularly
in gamma-rays, where rapid, energetic flaring events have been
detected in many sources. The production of large gamma-ray flares in
blazars appears to be closely connected with activity in the
parsec-scale jets. \cite{LV03} established a connection between flare
activity at gamma-ray and mm-wave energies, and \cite{JMMA01} have
reported that gamma-ray flares are often associated with the
appearance of new, superluminal features in the parsec scale jet.

The MOJAVE sample offers a new means of comparing the parsec-scale
jets of EGRET and non-EGRET sources, since it is selected on the basis
of compact radio flux density and is not biased with respect to
gamma-ray flux.  It contains 26 AGN considered to be
``high-probability'' EGRET identifications by \cite{MHR01} and
Sowerds-Emmerd et al. (2003, 2004), as well as 12 ``probable''
identifications.  We have divided the MOJAVE sample according to these
EGRET classifications (listed in Table 1) and performed standard
statistical tests on various properties to see if they originate from
the same parent population. For the purposes of our tests we have
grouped the ``probable'' identifications into the EGRET blazar class.

\begin{figure}
   \epsscale{.9}
   \plotone{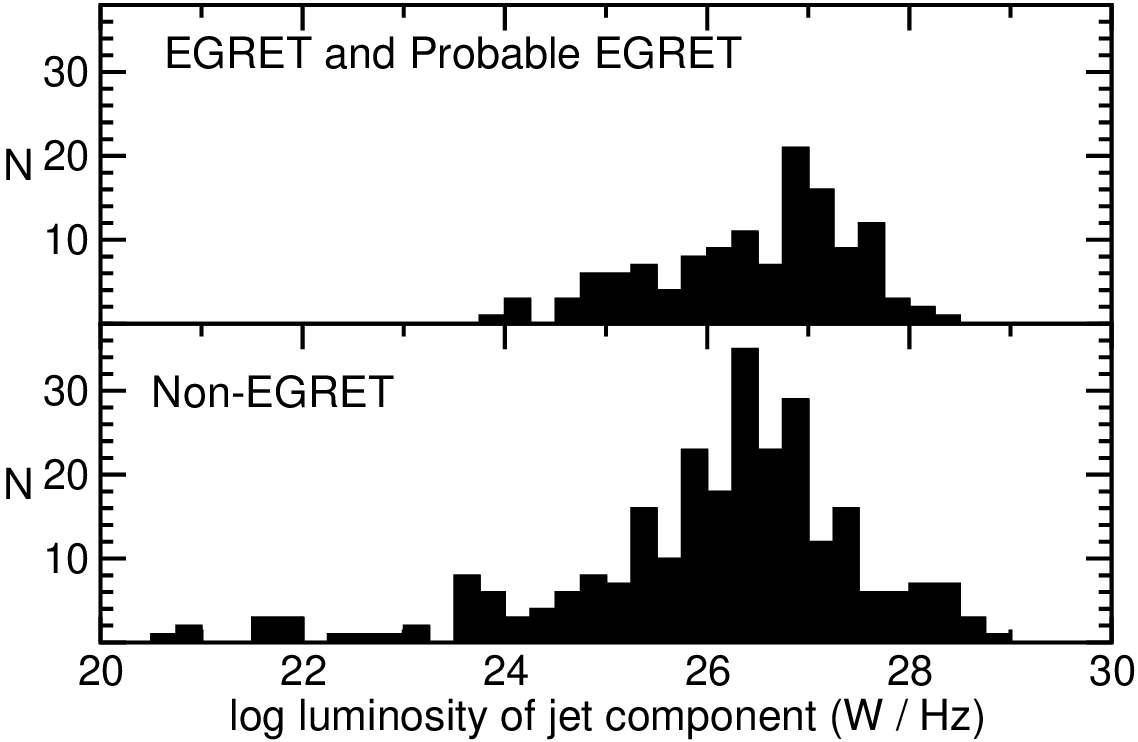}
   \caption{\label{f:EGRETLcpt} Distribution of luminosity at 15 GHz for
   individual jet components in gamma-ray (EGRET-detected) AGN (top panel),
   and non-EGRET AGN (lower panel) in the MOJAVE sample. }
   \end{figure}

We find that the EGRET and non-EGRET blazars have statistically
indistinguishable optical magnitude and redshift distributions. There
is a marginal difference in the total 15 GHz VLBA luminosity
distributions (94\% confidence), which increases to 99\% confidence if
the probable-EGRET sources are counted instead as non-EGRET
blazars. There is also a marginal indication (93\% confidence) of a
difference in the core luminosity distributions. Clear differences are
present in the parsec-scale jet features. In Figure~\ref{f:EGRETLcpt}
we show the 15 GHz luminosity distribution of all the Gaussian
components fitted to the MOJAVE sample, subdivided according to
gamma-ray classification. The median component luminosity in gamma-ray
blazars jets is twice that of the non-EGRET blazars, and the K-S test
gives only a 0.3\% chance that the luminosities are drawn from the
same parent distribution.

\begin{figure} 
	\plotone{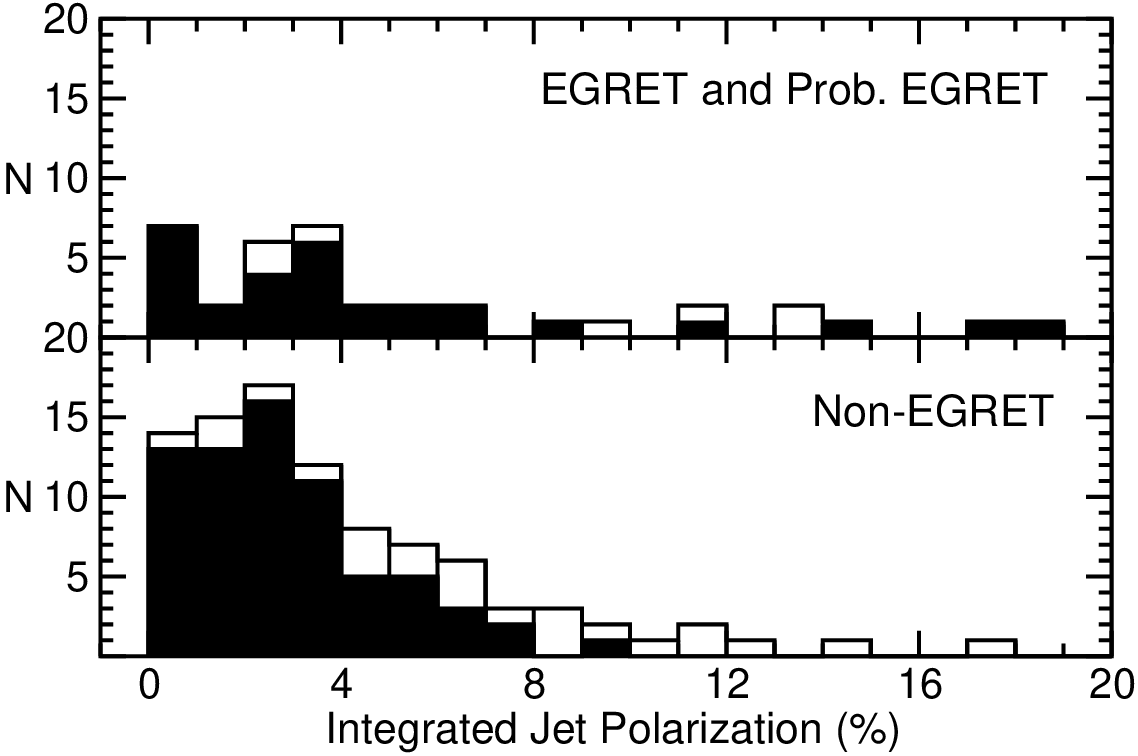} \caption{\label{f:EGRET_mjet}
	Distribution of integrated fractional polarization for the
	jets of EGRET (top panel) and non-EGRET (lower panel) AGN in
	the MOJAVE sample.}  \end{figure}

\begin{figure} 
	\plotone{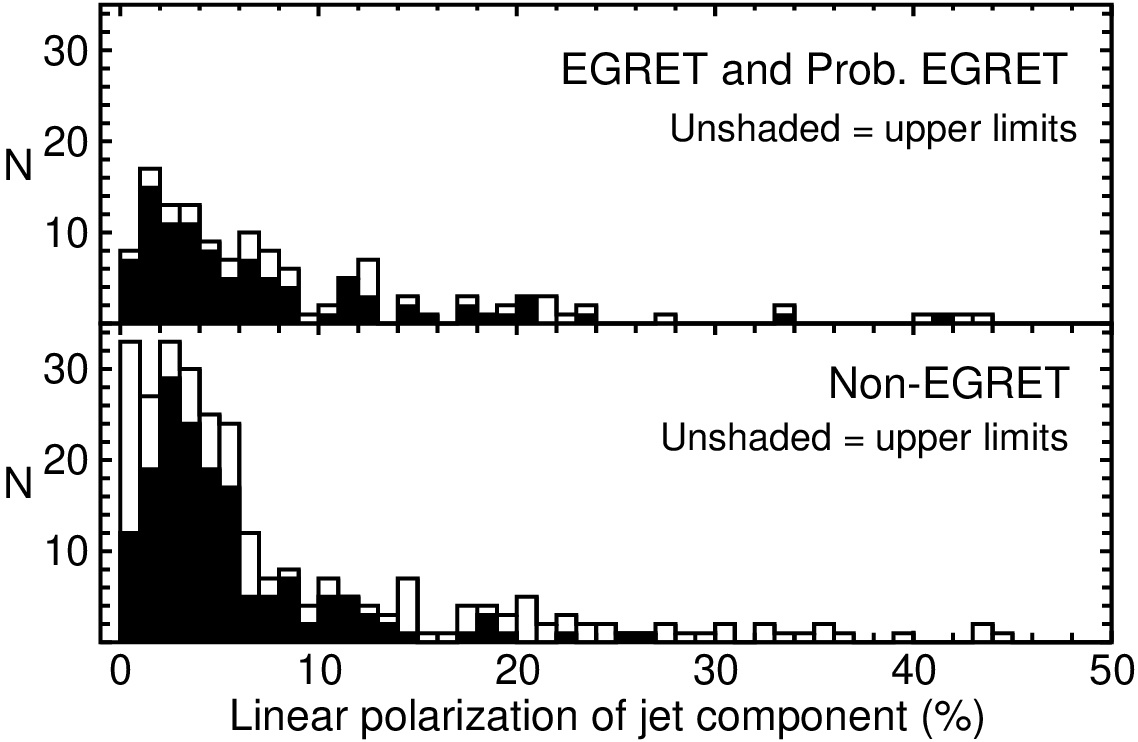} \caption{\label{f:EGRETmcpt}
	Distribution of fractional polarization for individual
	components in the jets of EGRET (top panel) and non-EGRET
	(lower panel) AGN in the MOJAVE sample.  The two distributions
	differ at the 99.9\% confidence level according to a Kendall's
	tau test on censored data.}  \end{figure}

The jets in EGRET sources also have statistically higher fractional
polarization levels, with a mean integrated jet polarization of 3.6\%
compared to 2.3\% for non-EGRET sources.  A survival analysis
comparison of the jet polarization distributions
(Fig.~\ref{f:EGRET_mjet}) indicates a difference at the 95\%
confidence level (99.99\% if the probable EGRET sources are
reclassified as non-EGRET). This is also the case for the polarization
distributions of individual jet components
(Fig.~\ref{f:EGRETmcpt}). Considering the distributions of
polarization detections alone, the differences in the two classes are
not large, however, the distribution of upper limits in
Fig.~\ref{f:EGRETmcpt} indicates that the gamma-ray blazar jets are
more highly polarized overall.  Unlike the case of BL Lacs versus
quasars, we see no tendency for EGRET jets to be more polarized at a
given distance from the core than non-EGRET jets.

These differences are perhaps unexpected, since they occur tens to
hundreds of parsecs downstream of the region where most models suggest
the gamma-rays are produced (near the base of the jet). It is possible
that the correlated jet and gamma-ray properties of the MOJAVE blazars
are not intrinsic to the jets, but are instead induced by Doppler
effects. EGRET blazars display a wider range of superluminal speeds
\citep{KL04}, which could imply that they have higher
Lorentz factors and/or jet axes aligned more closely with the line of
the sight. This would boost the apparent luminosity of their
finite-lifetime, steep-spectrum jet features more than their
non-gamma-ray counterparts. To reproduce the factor of two seen in the
MOJAVE sample, a difference in viewing angle of only 3 degrees is
needed, assuming a Lorentz factor of 10 and a spectral index of
$-0.7$. This number increases to approximately 10 degrees for a
Lorentz factor of 3. Such differences in viewing angle may have a
large effect on observed polarization detections, since
orthogonally-polarized regions may become blended via projection
effects, effectively canceling their polarized signal. Also, for high
Lorentz factor flows, a small change in viewing angle can translate
into a large change in aberrated angle in the jet rest frame, leading
to different levels of observed polarization \citep{HAA85}. It is not
clear how the observed jet electric vectors are affected in detail by
viewing angle changes, since this depends largely on the mechanism
that produces the polarization. If the polarization is from strong,
transverse shocks, the E-vectors should always align with the jet
flow. Unlike the case of BL Lacs versus quasars, we find no
differences in the electric vector offset distributions of the EGRET
versus non-EGRET sources, which argues against intrinsic magnetic
field differences.

Since the observed EGRET/non-EGRET polarization differences are
substantially different than those seen between BL Lacs and quasars,
more than one mechanism is probably at work.  A Fisher's exact test on
Table~\ref{t:optEGRET} shows no indication that the gamma-ray
properties of MOJAVE sources are related to their optical properties
(i.e., gamma-ray detections do not favor quasars over BL Lacs). BL
Lacs show strong indications of a dramatically different magnetic
field structure than quasars, perhaps as a consequence of lower jet
power.

Although the Doppler factor scenario appears to account for the
differences in gamma-ray blazar jet properties, it would also suggest
that these should extend to the core components. We only find a
marginal difference in the core luminosities of EGRET versus non-EGRET
blazars, and no differences in the core polarization properties. The
core luminosity may depend less strongly on Doppler factor because of
its flat spectral index and the fact that its emitting plasma is
likely to be continuously re-supplied (e.g., \citealt*{LB85}). But as
pointed out by \cite{L01}, the relation may not be not
straightforward, since it must take into consideration inverse-Compton
losses, finite lifetimes of unresolved components, and possible
deceleration of the flow. It is also possible that the high
variability level of the cores is masking any differences in our
single-epoch dataset.

As a further test of our hypothesis, we have divided the component
luminosities by the core luminosity in each source and applied
k-corrections using typical spectral indices of zero and $-0.7$ for
the core and jet, respectively. When comparing these ``normalized''
component luminosities in which the dependence on Doppler factor has
been reduced, we find no statistical differences between the
EGRET and non-EGRET sources, thereby supporting our scenario. 

In summary, the jet properties of gamma-ray blazars investigated here
can be reconciled by assuming that they have higher Doppler factors
than most blazars. This scenario can be tested more thoroughly using
more complete statistics on gamma-ray blazars observed by the GLAST
mission.

\section{Summary}

We have obtained first-epoch VLBA 15 GHz linear polarization images of
the first large complete AGN sample to be selected solely on the basis
of compact (milliarcsecond-scale) flux density.  This project,
entitled MOJAVE (Monitoring of Jets in Active Galactic Nuclei with
VLBA Experiments), is a complement to the VLBA 2 cm Survey of
\cite{KL04}.  Of the 133 AGN that satisfy our selection criteria, 88\%
are flat-spectrum blazars having relativistic jets pointed nearly
directly at us. The remainder consist of powerful, nearby radio
galaxies or peaked-spectrum radio sources. The sample contains 22
weak-lined BL Lacertae objects, and 38 AGN previously detected in
high-energy gamma-rays by EGRET. At least 94\% of the sample members
have a typical one-sided ``core-jet'' morphology, which consists of a
compact, bright core component located at the extreme end of a more
diffuse jet. The very small number of sources with visible
counter-jets reflects a strong Doppler favoritism toward jets viewed
end-on.


{\bf Core component properties:} The cores typically have brightness
temperatures exceeding $10^{11}$ K, and 37\% are completely
unresolved, with lower limit values ranging from $\sim 10^{12}$ to
$10^{14}$ K. The unresolved cores are more likely to show good
alignments between their electric polarization vectors and the
direction of the inner jet, perhaps as a result of a strong transverse
shock located near the base of the jet. The cores of BL Lacertae
objects show a particularly strong tendency toward these alignments,
and have higher polarization levels than the quasars. The majority of
cores in the sample have fractional polarization levels below 5 \%,
but the electric vector alignments suggest that this is not because of
Faraday effects. Faraday depolarization is likely present, however, in
the radio galaxy cores in our sample, none of which were detected in
polarized emission.

{\bf Jet properties:} We confirm the earlier findings of \cite{L01}
that the jets of BL Lacs are more polarized than quasars at a given
distance from the core. We also find that the trend of increasing
fractional polarization with distance down the jet in blazars found 
by \cite{L01} extends to projected distances of at least 100 parsecs
downstream.  This increase in magnetic field order is not accompanied
by an increase in electric vector alignment as found at 43 GHz by
\cite{LS00}, although we note that our observations have three times
poorer spatial region and probe regions much further downstream.  We
find many jet regions to have extremely high fractional polarization
(> 50\%), which indicates highly ordered magnetic fields in
optically thin plasma. The electric vectors of the polarized regions
are well-aligned with the jet ridge line in the case of BL Lac
objects, consistent with strong, transverse shocks.  The electric
vectors in quasar jets have a variety of orientations, however,
indicating either a different polarization mechanism, or a tendency
toward oblique shocks.

Our results imply that the magnetic field properties of parsec-scale
jets are closely tied to the broad line region in blazars. BL Lac
objects have been postulated to be the end-on versions of lower power
FR-I type radio galaxies \cite{UP95}, and in such jets shocks may form
more easily, leading to different polarization structures than those
found in high-power quasars. 

{\bf Gamma-Ray Blazars:} We have examined the properties of the 35
blazars in MOJAVE that have been identified as likely candidates to
EGRET gamma-ray sources. The jet features in these sources are
typically twice as luminous as those in non-EGRET source, and are more
likely to be linearly polarized. Since these differences occur tens to
hundreds of parsecs downstream of where the gamma-rays are thought to
be produced, the most likely explanation is that gamma-ray AGN have
slightly higher Doppler factors than typical blazars, in agreement
with the conclusions of \cite{KL04}.  

\acknowledgements
 The authors wish to acknowledge the contributions of the other
 members of the MOJAVE project team: Hugh and Margo Aller, Tigran
 Arshakian, Marshall Cohen, Matthias Kadler, Ken Kellermann, Yuri
 Kovalev, Andrei Lobanov, Eduardo Ros, Rene Vermeulen, and Tony
 Zensus.

 This research was supported by NSF grant 0406923-AST and made use of
 the following resources: 
 
 The NASA/IPAC Extragalactic Database (NED), which is operated by the
 Jet Propulsion Laboratory, California Institute of Technology, under
 contract with the National Aeronautics and Space Administration.

 The VizieR database of astronomical catalogs \citep{OBM00}. 
 
 The University of Michigan Radio Astronomy Observatory, which
 is supported by the National Science Foundation and by funds from the
 University of Michigan.

\mbox{RATAN--600} observations were partly
supported by the Russian State Program ``Astronomy'' and the Russian
Ministry of Education and Science, the NASA JURRISS Program grant
W-19611, and the Russian Foundation for Basic Research grant
01--02--16812.

\begin{deluxetable}{llrrlclcrc} 
\tablecolumns{10} 
\tabletypesize{\scriptsize} 
\tablewidth{0pt}  
\tablecaption{\label{gentable}The Complete MOJAVE Sample}  
\tablehead{\colhead{IAU} & \colhead {} &  
\colhead{R.A.} &\colhead{Dec.} & \colhead{} & 
\colhead{Opt.} & &  
\colhead{Radio}  &  \colhead{$S_{VLBI}$} &\colhead{$\gamma$-ray} \\ 
\colhead{Name} & \colhead {Alias} &  
\colhead{(J2000)} & \colhead{(J2000)} &\colhead{z} &\colhead{Class} & \colhead{V} &  
\colhead{Spectrum}  & \colhead{(Jy)}\ &  \colhead{ID} \\ 
\colhead{(1)} & \colhead{(2)} & \colhead{(3)} & \colhead{(4)} &  
\colhead{(5)} & \colhead{(6)} & \colhead{(7)} & \colhead{(8)} & \colhead{(9)}& \colhead{(10)} } 
\startdata 
0003$-$066  & NRAO 005  & 00 06 13.8928  & $-$06 23 35.3353 & 0.347 & B& 18.5 & Flat &  2.83\phn & \n    \\
0007+106  & III Zw 2  & 00 10 31.0058  & $+$10 58 29.5041 & 0.0893 & G& 15.4 & Flat &  3.00* & \n    \\
0016+731  &   & 00 19 45.7864  & $+$73 27 30.0174 & 1.781 & Q& 19.0 & Flat &  2.40* & \n    \\
0048$-$097  &   & 00 50 41.3173  & $-$09 29 05.2102 & \n & B& 17.4 & Flat &  2.70* & \n    \\
0059+581  &   & 01 02 45.7623  & $+$58 24 11.1366 & \n & U& \n & Flat &  3.31\phn & \n    \\
0106+013  &   & 01 08 38.7710  & $+$01 35 00.3171 & 2.107 & Q& 18.4 & Flat  &  2.97\phn & \n    \\
0109+224  &   & 01 12 05.8247  & $+$22 44 38.7861 & \n & B& 15.7 & Flat  &  1.65* & \n    \\
0119+115  &   & 01 21 41.5950  & $+$11 49 50.4131 & 0.570 & Q& 19.7 & Flat  &  2.00* & \n    \\
\enddata 
\tablecomments{Columns are as follows: (1) IAU Name (B1950.0); (2) Other Name; (3) Right Ascension (J2000); (4) Declination (J2000); (5) Redshift from \cite{VCV03} (the redshift of 0202+149 is from \citealt*{SRK96}, that of 0727$-$115 is from \citealt*{Z02}, that of 0742+103 is from \citealt*{BPB03}, that of 0754+100 is from \citealt*{CFK03}, that of 0808+019 is from \citealt*{JWS02}, and that of 2209+236 is from \citealt{SRM03});  (6) Optical classification according to the \cite{VCV03} catalog, where Q = quasar, B = BL Lac object, G = active galaxy, and U = unidentified; (7) Optical V magnitude from  \cite{VCV03} catalog; (8) Description of radio spectrum from \cite{KL04}; (9) Maximum 2 cm total cleaned VLBA flux density at any epoch during 1994--2003. Values with asterisks are inferred from single-dish observations (see \S \ref{sampselect}); (10) EGRET gamma-ray source identification according to \cite{MHR01}, \cite{SRM03} and \cite{SRM04}, where Y = highly probable identification, P = probable identification.}

\tablenotetext{a}{Source classified as a quasar in the \cite{VCV03} catalog.}
\tablenotetext{b}{Source classified as a possible BL Lac object in the \cite{VCV03} catalog.}
\tablenotetext{c}{Source classified as a galaxy in the \cite{VCV03} catalog.}

\end{deluxetable} 
\begin{deluxetable}{lccccc} 
\tablecolumns{6} 
\tabletypesize{\scriptsize} 
\tablewidth{0pt}  
\tablecaption{\label{t:optEGRET}Optical and EGRET Gamma-Ray Classifications of
the MOJAVE Sample}  
\tablehead{\colhead{} & \colhead {Quasars} &  
\colhead{BL Lac Objects} &\colhead{Radio Galaxies} & \colhead{Unidentified} & 
\colhead{Total} } 
\startdata 
EGRET & 20 & 6 & 0 & 0 & 26 \\
Probable EGRET & 12 & 0 & 0 & 0 & 12 \\
Non-EGRET & 62 & 16 & 8 & 9 & 95 \\
Total & 94 & 22 & 8 & 9 & 133 \\
\enddata

\end{deluxetable} 
\begin{deluxetable}{llccrrrrrrrr} 
\tablecolumns{12} 
\tabletypesize{\scriptsize} 
\tablewidth{0pt}  
\tablecaption{\label{maptable}Summary of 15 GHz Image Parameters}  
\tablehead{\colhead{Source} & \colhead {Epoch} &  
\colhead{$\mathrm{B_{maj}}$} &\colhead{$\mathrm{B_{min}}$} & \colhead{$\mathrm{B_{pa}}$} &  
\colhead{$\mathrm{I_{tot}}$} & \colhead{$\mathrm{I_{peak}}$}&  \colhead{$\mathrm{P_{tot}}$} & \colhead{$\mathrm{EVPA_{tot}}$} & \colhead{$\mathrm{I_{base}}$} &\colhead{$\mathrm{P_{base}}$} &\colhead{Scale} \\ 
\colhead{(1)} & \colhead{(2)} & \colhead{(3)} & \colhead{(4)} &  
\colhead{(5)} & \colhead{(6)} & \colhead{(7)} & \colhead{(8)} & \colhead{(9)}& \colhead{(10)} & \colhead{(11)} & \colhead{(12)} } 
\startdata 
0003$-$066  & 2003 Feb 6 &  1.32 &  0.54 & $-$5 &   2.83 &   1.83 &  101 & 17 &  1.7 &  1.7 & 4.9  \\ 
0007+106  & 2004 Apr 10 &  1.16 &  0.51 & $-$6 &   1.38 &   1.35 &  2 & 10 &  0.8 &  0.7 & 1.7  \\ 
0016+731  & 2003 Aug 28 &  0.74 &  0.64 & $-$87 &   1.07 &   1.01 &  9 & 108 &  0.7 &  1.7 & 8.4  \\ 
0048$-$097  & 2003 Aug 28 &  1.30 &  0.49 & $-$8 &   0.74 &   0.70 &  30 & 28 &  0.8 &  1.2 & \nodata  \\ 
0059+581  & 2002 Nov 23 &  0.78 &  0.62 & $-$9 &   3.30 &   3.16 &  63 & 11 &  1.1 &  1.2 & \nodata  \\ 
0106+013  & 2003 Mar 29 &  1.20 &  0.51 & $-$5 &   2.86 &   2.13 &  35 & 178 &  0.8 &  2.3 & 8.3  \\ 
0109+224  & 2002 Jun 15 &  1.05 &  0.56 & $-$6 &   1.01 &   0.93 &  43 & 86 &  0.5 &  0.9 & \nodata  \\ 
0119+115  & 2003 Mar 1 &  0.96 &  0.47 & $-$2 &   1.25 &   0.87 &  38 & 179 &  0.8 &  1.1 & 6.5  \\
\enddata 
\tablecomments{Columns are as follows: (1) IAU Name (B1950.0); (2) Date of VLBA observation; (3) FWHM major axis of restoring beam (milliarcseconds); (4) FWHM minor axis of restoring beam (milliarcseconds); (5) Position angle of major axis of restoring beam; (6) Total I flux density (Jy); (7) Peak flux density in the image (Jy per beam);  (8) Total linearly polarized flux density (mJy); (9) Integrated electric vector position angle (deg.); (10) Lowest I contour in image (mJy per beam) (11) Lowest P contour in image (mJy per beam). (12) Linear scale in parsecs per milliarcsec. }

\end{deluxetable} 
\begin{deluxetable}{lcrrccrrrrr} 
\tablecolumns{11} 
\tabletypesize{\scriptsize} 
\tablewidth{0pt}  
\tablecaption{\label{coretable}Core Component and Integrated Jet Properties }  
\tablehead{\colhead{Source} &\colhead{r} & \colhead{P.A.} & \colhead {$I_\mathrm{c}$} &  
\colhead{$\mathrm{Maj.}$} &\colhead{$\mathrm{Ratio}$} & \colhead{$\mathrm{P.A.}$} &  
\colhead{$\mathrm{T_{b,c}}$} & \colhead{$m_\mathrm{c}$}&  \colhead{$\mathrm{EVPA_{c}}$}& \colhead{$m_\mathrm{jet}$}   \\ 
\colhead{(1)} & \colhead{(2)} & \colhead{(3)} & \colhead{(4)} &  
\colhead{(5)} & \colhead{(6)} & \colhead{(7)} & \colhead{(8)} & \colhead{(9)}& \colhead{(10)}& \colhead{(11)}  } 
\startdata 
0003$-$066 & \n & \n  &   2.47 & 0.93 & 0.26 & 1 &  10.8 &  4.3 & 195  & 3.5    \\ 
0007+106 & \n & \n  &   0.94 & \n & \n & \n & $>$ 12.9 & $<$ 0.06 & \n  &$<$ 14.6    \\ 
0016+731 & \n & \n  &   1.03 & 0.13 & 0.60 & 289 &  11.7 &  1.0 & 112  &$<$ 7.4    \\ 
0048$-$097 & \n & \n  &   0.73 & 0.21 & 0.34 & 12 &  11.4 &  4.2 & 209  &$<$ 17.0    \\ 
0059+581 & \n & \n  &   3.25 & 0.13 & 0.85 & 8 &  12.1 &  2.0 & 191  &$<$ 5.9    \\ 
0106+013 & \n & \n  &   1.84 & 0.28 & 0.39 & 3 &  11.5 &  1.5 & 185  & 2.4    \\ 
0109+224 & \n & \n  &   0.86 & \n & \n & \n & $>$ 12.9 &  4.6 & 87  & 4.8    \\ 
0119+115 & \n & \n  &   0.90 & 0.30 & 0.13 & 26 &  11.6 &  5.6 & 0  & 5.9    \\
\enddata 
\tablecomments{Columns are as follows: (1) IAU Name (B1950.0); (2) Core position offset from the map center; (3) Core position angle with respect to the map center; (4) Fitted Gaussian core flux density; (5) FWHM major axis of fitted core Gaussian (milliarcseconds); (6) Axial ratio of fitted core Gaussian; (7) Major axis position angle of fitted core Gaussian; (8) Log of core brightness temperature (K); (9) Fractional polarization at core position;  (10) Electric vector position angle at core position;  (11) Integrated fractional polarization of jet (per cent).}

\end{deluxetable} 
\begin{deluxetable}{lcrrccrrr} 
\tablecolumns{9} 
\tabletypesize{\scriptsize} 
\tablewidth{0pt}  
\tablecaption{\label{cpttable}Jet  Component Properties }  
\tablehead{\colhead{Source}  & \colhead{r} & \colhead {P.A.} &  
\colhead{$\mathrm{Maj.}$} &\colhead{$\mathrm{Ratio}$} & \colhead{$\mathrm{P.A.}$} &  
 \colhead{I} & \colhead{$m_\mathrm{}$}&  \colhead{$\mathrm{EVPA_{}}$} \\ 
\colhead{(1)} & \colhead{(2)} & \colhead{(3)} & \colhead{(4)} &  
\colhead{(5)} & \colhead{(6)} & \colhead{(7)} & \colhead{(8)} & \colhead{(9)}  } 
\startdata 
0003$-$066 & 0.89  & 260 & \n & \n &  \n  &79 &   11.1 & 42  \\ 
 & 1.96  & 304 & \n & \n &  \n  &58 &   18.8 & 16  \\ 
 & 6.63  & 281 & 1.61 & 0.62 &  359  &216 &   11.6 & 97  \\ 
0007+106 & 0.09  & 274 & \n & \n &  \n  &439 &  $<$ 0.06 & \n  \\ 
0016+731 & 1.24  & 136 & 1.27 & 0.58 &  310  &36 &  $<$ 14.5 & \n  \\ 
0048$-$097 & 1.25  & 359 & \n & \n &  \n  &9 &   4.7 & 208  \\ 
0059+581 & 1.43  & 235 & 2.63 & 0.27 &  72  &39 &  $<$ 20.4 & \n  \\ 
0106+013 & 0.30  & 235 & \n & \n &  \n  &506 &   2.6 & 201  \\
\enddata 
\tablecomments{Columns are as follows: (1) IAU Name (B1950.0);  (2) Distance from fitted core position (mas); (3) Position angle of component w.r.t. core position; (4) Major axis of fitted Gaussian (milliarcseconds); (5) Axial ratio of fitted Gaussian; (6) Position angle of major axis of fitted  Gaussian; (7) log of brightness temperature (K);(8) Fitted component flux density (mJy); (9) Fractional polarization at component position;  (10) Electric vector position angle at component position.}

\end{deluxetable} 

\end{document}